\documentclass[aps,prb,preprint,showpacs]{revtex4}
\usepackage{natbib}
\usepackage{epsfig}
\usepackage{color}

\begin{document}

\title{Pentagon chain with spin orbit interactions: exact many-body ground
states in the interacting case}
\author{Zsolt~Gulacsi} 
\affiliation{Department of Theoretical Physics, University of
Debrecen, H-4010 Debrecen, Bem ter 18/B, Hungary}
\date{\today }

\begin{abstract}
  Based on a positive semidefinite operator technique, exact ground states are
  deduced for the non-integrable conducting polymers possessing pentagon type of
  unit cell. The study is done in the presence of many-body spin-orbit
  interaction (SOI), local and nearest neighbor Coulomb repulsion (NNCR) and
  presence of external $E$ electric and $B$ magnetic fields, such that the
  effects of $B$ on both orbital and spin degrees of freedom is considered. The
  SOI, NNCR, and presented external field configurations presence in exact
  conducting polymer ground states is a novelty, so the development of the
  technique for the treatment possibility  of such strongly correlated cases
  is presented in details. The deduced ground states show a broad spectrum of
  physical characteristics ranging from charge density waves, metal-insulator
  transitions, to interesting external field driven effects as e.g.
  modification possibility of a static charge distribution by a static external
  magnetic field.  
\end{abstract}

\maketitle

\section{Introduction}

Given by their multifunctional characteristics as simplistic synthesis,
environmental stability, beneficial electronic, mechanic and optical properties,
low cost and weight, or biocompatibility, the conducting polymers are largely
used on a broad spectrum of applications in advanced technology as: 
electrochemistry applications, electrochemical sensing, energy storage,
supercapacitors, batteries, sensors, fuel or solar cells, or drug delivery in
organism, etc \cite{Intr1}.

If many-body spin-orbit interaction (SOI) is present in these systems, their
applicability is enlarged, and new properties come in. E.g. applications in
spintronics become possible \cite{Intr2}; the possibilities to relax the
rigid mathematical conditions leading to flat bands become available
\cite{Intr3}, allowing the emergence of different ordered phases by small
perturbations; paves the way for spin orbitronics in plastic materials, and
leads even to inverse spin Hall effect opening the doors for topological
behavior \cite{Intr2}. It must be underlined, that even if the SOI coupling
($\lambda$) sometimes is small in some organic materials, it can be
substantially enhanced (even continuously tuned) by external fields
\cite{Intr3}, pulsed ferromagnetic resonance \cite{Intr2}, or introduction of
intrachain atoms that considerably enhance the spin-orbit coupling (e.g.
platinum) \cite{Intr4}.

It should be also noted, that the importance of SOI is stressed as well by
the fact, that even in the case when the spin-orbit coupling is small, i.e.
$\lambda << 1$, its effect is major, because it breaks the spin-projection
double degeneracy of each band \cite{Intr5}. Furthermore, usually these systems
are interacting, hence the leading term of the Coulomb interaction in many-body
systems, the on-site Coulomb repulsion $U > 0$, attains even high values in
organic materials \cite{Intr6}. In these conditions also the nearest
neighbor Coulomb repulsion $V > 0$ satisfying $V < U$ influences the physical
behavior. So in the description of these systems, the two extreme
characteristic parameters ($\lambda, U$) satisfy the $U >> \lambda$ relation.
Consequently the perturbative treatment becomes questionable in both low and
high coupling constant limits, enforcing
special treatment for obtaining exact results for a good quality description.
The ``special treatment'' is accentuated here because these systems are
usually non-integrable, so Bethe-anzats type of treatment in such cases is
inapplicable. In these condition exact results for conducting polymers in the
presence of SOI till present are not known. The aim of this paper is to break
this state of facts, and to present the first exact results for conducting
polymers in the presented conditions ($\lambda$, $V$ and $U$ present with
$U >> \lambda$ condition) in many-body, strongly correlated case. I note that
one particle type exact solutions in the presence of SOI have been already
published in the bosonic 1D situations \cite{Intr7}, but many-body interacting
fermionic exact solutions in the presence of SOI for conducting polymers are
not known.
 
An often studied representative of conducting polymers is the polyaminotriazole
type of chain with pentagon unit cell, as shown in Fig.1, which  will be
analyzed also here. The procedure we use is based on positive semidefinite
operator properties, for which you can find detailed description e.g. Ref.
\cite{Intr8} or even review papers as Ref.\cite{Intr9}. The procedure, which
works for non-integrable systems as well, in principle looks as follows: First
the Hamiltonian of the system is transformed in exact terms in a positive
semidefinite form $\hat H = \sum_ {\nu} \hat P_{\nu} +C$, where $C$ is a scalar,
and $\hat P_{\nu}$ are positive semidefinite operators. The transformation of
the Hamiltonian connects the parameters of the Hamiltonian to the parameters
of the $\hat P_{\nu}$ operators via a nonlinear system of matching equations,
which must be solved in order to complete the transformation process.
The positive semidefinite operators by definition satisfy the
$\langle \Psi| \hat P_{\nu}|\Psi \rangle \geq 0$ relation, hence their minimum
possible eigenvalue is zero. This property underlines the advantages, and force
of attraction of the positive semidefinite operator procedure: instead to try
to deduce an arbitrary valued ground state, we can concentrate to the deduction
of a ground state that has a well defined position, which often represents a
much doable task. Since we may think, that this job requires to a priory know,
or guess the value of the $C$ scalar from the transformation, I must underline
that this is not true. $C$ is connected in several equations to the parameters
of the operators $\hat P_{\nu}$, and to the finally deduced ground state
expression (which is also connected to the parameters of  $\hat P_{\nu}$), so
$C$ becomes to be known only at the end of the calculation.

Consequently
$\hat H' = \hat H -C$ has the minimum possible eigenvalue zero corresponding to
the ground state eigenfunction $|\Psi_0\rangle$. This last, is deduced from
$(\sum_ {\nu} \hat P_{\nu})|\Psi_0\rangle=0$, using elevated techniques, see
e.g. Ref.\cite{Intr8,Intr9}. The procedure provides valuable results even in
three dimensions \cite{Intr10}, or two dimensional disordered
systems \cite{Intr11}.
\begin{figure}[h]
\includegraphics[height=5cm, width=6cm]{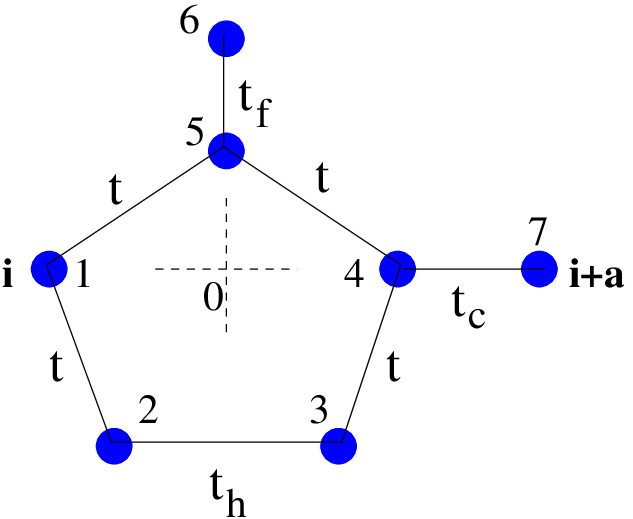}
\caption{The pentagon unit cell. The numbers denote the n value, the in cell
  numbering of sites. The origin of the system of coordinates from where one
  analyses the cell is denoted by 0. The hopping matrix elements $t,t_c,t_h,
  t_f$are indicated on bonds, ${\bf i}$ denotes the lattice site to which the
  cell corresponds, and ${\bf a}$ is the lattice constant.}
\end{figure}

Concerning the technique used, it must be underlined \cite{Intr9} that contrary
to the early stage of the method of the positive semidefinite operators where
in the first step the ground state has been constructed, and the Hamiltonian
adjusted to it (usually by extension terms in the Hamiltonian), the here used
procedure starts from a given Hamiltonian, and deduce coherently the ground state
corresponding to it without to know it a priory, and without to use extension
terms. Concerning the pentagon
chains, till now only nearest neighbor hoppings and the Hubbard interactions
have been considered. The novelty in the present paper is that besides the
Hamiltonian terms mentioned above, in deducing the exact ground states, we
also consider SOI, nearest-neighbor Coulomb repulsion, and presence of external
fields, from which the effect of the external magnetic field is considered both
in its effect on the orbital motion of carriers (by Peierls phase factors), and
its effect on the spin degrees of freedom (by the Zeeman term).
In these conditions the paper focuses on the development of the technique such
to construct a procedure which allows the treatment of the system in the
presented circumstances. Hence one concentrates 
on the transformation of the Hamiltonian in positive semidefinite form,
solution strategy of the matching equations, and construction process of the
exact ground states, items, that for a coherent presentation need a given, and
not negligible room. Hence, the presentation of the physical properties of the
deduced ground states and phases is only sketched (I also note that this job
needs extensive further supplementary study), the detailed presentation of
the physical
properties of the deduced exact ground states is left to a future publication.  
Nevertheless, the presented deduced physical properties underline a broad
spectrum of interesting characteristics as e.g. emerging charge density wave
phases, modification possibility of a static charge distribution by an
external static magnetic field, switching possibilities between insulating and
conducting phases by external magnetic fields, creation of effective flat bands
provided not by the bare band structure, but by the two-particle local or
non-local Coulomb interactions, peculiar ferromagnetic states, etc.

The remaining part of the paper is structured as follows: Section II presents
the studied system, the transformation of the Hamiltonian in positive
semidefinite form, and the solutions of the matching equations in the presence
and absence of external fields. Section III presents the deduction procedure of
the exact ground states, and the obtained ground states themselves. Section IV
presents the summary and conclusions, while the attached six Appendices at the
end of the paper contain the mathematical details.

\section{The system analyzed}

\subsection{The case of the zero external magnetic field}

The Hamilton of the system is $\hat H = \hat H_K + \hat H_U + \hat H_V$,
where one has
\begin{eqnarray}
\hat H_K &=& \sum_i \sum_{\sigma,\sigma'} [t^{\sigma,\sigma'}(\hat c^{\dagger}_{i,1,\sigma}
\hat c_{i,5,\sigma'} + \hat c^{\dagger}_{i,2,\sigma} \hat c_{i,1,\sigma'} +
\hat c^{\dagger}_{i,4,\sigma} \hat c_{i,3,\sigma'} + \hat c^{\dagger}_{i,5,\sigma} 
\hat c_{i,4,\sigma'}) + t_h^{\sigma,\sigma'} \hat c^{\dagger}_{i,3,\sigma}
\hat c_{i,2,\sigma'}
\nonumber\\
&+& t_f^{\sigma,\sigma'} \hat c^{\dagger}_{i,6,\sigma} \hat c_{i,5,\sigma'} 
+ t_c^{\sigma,\sigma'} \hat c^{\dagger}_{i+a,7,\sigma} \hat c_{i,4,\sigma'} + h.c.] +
\hat H_{\epsilon}, \quad
\hat H_{\epsilon} =
\sum_{i,n}\epsilon^{\sigma,\sigma'} \hat c^{\dagger}_{i,n,\sigma} \hat c_{i,n,\sigma'},
\nonumber\\
\hat H_U &=& \sum_i \sum_{n=1,2,..6} U_n \hat n_{i,n}^{\uparrow}
\hat n_{i,n}^{\downarrow}, \quad
\hat H_V = V \sum_{<i,j>} \sum_{<n,n'>} \hat n_{i,n} \hat n_{j,n'}
\label{Ez1}
\end{eqnarray}
In Fig.1 one presents the unit cell of the system containing the in cell
numbering of sites ($n$) in between 1 and 6. The hopping matrix elements,
and the on-site one particle potentials have $(\sigma,\sigma')$ spin projection
indices in order to allow the presence of the many-body SOI. To match better
the experimental situations, the antenna (describing the upper connected
atoms), i.e. bond 5-6, the lower flank (bond 2-3), and the inter-cell
connection (bond 4-7), are such taken to allow different hopping matrix
elements, again for allowing comparison to different experimental realizations
of the pentagon chain. The $\hat H_{\epsilon}$ term collects the on-site one
particle potential terms. The $U_n > 0$ coefficients are representing the on
site Coulomb repulsions (Hubbard interaction), while $V \geq 0$ represents the
nearest-neighbor Coulomb repulsion.

In order to transform in exact terms $\hat H$ in a positive semidefinite form,
one introduces ten block operators $\hat B_{i,z,\sigma}$, $z=1,2,...5$ as follows:
\begin{eqnarray}
&&\hat B_{i,1,\uparrow}= a_{1,1} \hat c_{i,1,\uparrow} +a_{1,2} \hat c_{i,2,\uparrow} +
a_{1,5} \hat c_{i,5,\uparrow} + b_{1,1} \hat c_{i,1,\downarrow},
\nonumber\\
&&\hat B_{i,1,\downarrow}= a_{1,1} \hat c_{i,1,\downarrow} +a_{1,2} \hat c_{i,2,\downarrow}
+ a_{1,5} \hat c_{i,5,\downarrow} + b'_{1,1} \hat c_{i,1,\uparrow},
\nonumber\\
&&\hat B_{i,2,\uparrow}= a_{2,2} \hat c_{i,2,\uparrow} +a_{2,3} \hat c_{i,3,\uparrow} +
a_{2,5} \hat c_{i,5,\uparrow},
\nonumber\\
&&\hat B_{i,2,\downarrow}= a_{2,2} \hat c_{i,2,\downarrow} +a_{2,3} \hat c_{i,3,\downarrow}
+ a_{2,5} \hat c_{i,5,\downarrow},
\nonumber\\
&&\hat B_{i,3,\uparrow}= a_{3,3} \hat c_{i,3,\uparrow} +a_{3,4} \hat c_{i,4,\uparrow} +
a_{3,5} \hat c_{i,5,\uparrow} + b_{3,4} \hat c_{i,4,\downarrow},
\nonumber\\
&&\hat B_{i,3,\downarrow}= a_{3,3} \hat c_{i,3,\downarrow} +a_{3,4} \hat c_{i,4,\downarrow}
+ a_{3,5} \hat c_{i,5,\downarrow} + b'_{3,4} \hat c_{i,4,\uparrow},
\nonumber\\
&&\hat B_{i,4,\uparrow}= a_{4,5} \hat c_{i,5,\uparrow} +a_{4,6} \hat c_{i,6,\uparrow},
\nonumber\\
&&\hat B_{i,4,\downarrow}= a_{4,5} \hat c_{i,5,\downarrow} +a_{4,6} \hat c_{i,6,\downarrow},
\nonumber\\
&&\hat B_{i,5,\uparrow}= a_{5,4} \hat c_{i,4,\uparrow} +a_{5,7} \hat c_{i+a,7,\uparrow} +
b_{5,4} \hat c_{i,4,\downarrow} + b_{5,7} \hat c_{i+a,7,\downarrow},
\nonumber\\
&&\hat B_{i,5,\downarrow}= a_{5,4} \hat c_{i,4,\downarrow} +a_{5,7}
\hat c_{i+a,7,\downarrow} +
b'_{5,4} \hat c_{i,4,\uparrow} + b'_{5,7} \hat c_{i+a,7,\uparrow}.
\label{Ez2}
\end{eqnarray}
Here $i$ represents the lattice site, $z$ denotes the block operator index
($z=1,2,...5$), so $(z,\sigma)$ has 10 different values, while
$\sigma$ represents the spin projection.
The coefficients $a_{z,n}$ and $b_{z,n}$ called
block operator parameters, are indexed by the block operator index $z$, and
$n$, the on cell position on which the operator following the coefficient acts.
The coefficients $a_{z,n}$ are present in terms whose spin projection is equal to
the spin projection of the block operator, while the $b_{z,n}$ coefficients
denote terms whose spin projection is opposite to the spin projection of the
block operator. One has totally 21 block operator parameters ($a_{z,n}$ values:
13; and $b_{z,n}$ values: 8), which at the moment are unknown, but can be
determined from the transformation of the starting Hamiltonian (\ref{Ez1}), to
the positive semidefinite Hamiltonian
\begin{eqnarray}
\hat H = \sum_{i,z,\sigma} \hat B^{\dagger}_{i,z,\sigma} \hat B_{i,z,\sigma} +
\hat H_U + \hat H_V.
\label{Ez3}
\end{eqnarray}

At this step I must note that the presented expression of the block operators
in (\ref{Ez2}) has been optimized for the analyzed problem. This optimization
has in fact two main aspects, namely: i) Because of the translational symmetry,
the block operator parameters are the same in each unit cell, i.e. are $i$
independent. ii) The opposite spin components in the block operators are
optimized in order to minimally describe the requirements raised by the
presence of SOI which leads in fact to spin-flip hoppings. Given by this, we
have much less $b_{z,n}$ coefficients in block operators than $a_{z,n}$
coefficients, which simplifies and helps the mathematical treatment.

If (\ref{Ez3}) is the same relation as (\ref{Ez1}), it must be a relationship
in between the block operator parameters and the starting parameters of the
Hamiltonian (\ref{Ez1}). These relations called the matching equations are
presented in the following subsection.

\subsection{Matching equations in zero external magnetic field.}

The matching equations are obtained by effectuating the $ \hat B^{\dagger}_{i,z,
 \sigma} \hat B_{i,z,\sigma}$ product together with the $\sum_{i,z,\sigma}$ sum
operation in
(\ref{Ez3}), and equating each obtained term to the corresponding term in
(\ref{Ez1}). E.g. the coefficient of the operator $\hat c^{\dagger}_{i,1,\uparrow}
\hat c_{i,5,\uparrow}$ in (\ref{Ez3}) is $a^*_{1,1}a_{1,5}$, while in (\ref{Ez1})
is $t_{1,5}^{\uparrow,\uparrow}$, the relation being spin projection independent,
from where the third line of (\ref{Ez4}) follows, etc.

For the hoppings without spin-flip one obtains ten equations:
\begin{eqnarray}
&&0=t^{\uparrow,\uparrow}_{2,5}=t^{\downarrow,\downarrow}_{2,5}=
a^*_{1,2}a_{1,5}+a^*_{2,2}a_{2,5},
\nonumber\\
&&0=t^{\uparrow,\uparrow}_{3,5}=t^{\downarrow,\downarrow}_{3,5}=
a^*_{3,3}a_{3,5}+a^*_{2,3}a_{2,5},
\nonumber\\
&&t=t^{\uparrow,\uparrow}_{1,5}=t^{\downarrow,\downarrow}_{1,5}=
a^*_{1,1}a_{1,5},
\nonumber\\
&&t=t^{\uparrow,\uparrow}_{2,1}=t^{\downarrow,\downarrow}_{2,1}=
a^*_{1,2}a_{1,1},
\nonumber\\
&&t=t^{\uparrow,\uparrow}_{4,3}=t^{\downarrow,\downarrow}_{4,3}=
a^*_{3,4}a_{3,3},
\nonumber\\
&&t=t^{\uparrow,\uparrow}_{5,4}=t^{\downarrow,\downarrow}_{5,4}=
a^*_{3,5}a_{3,4},
\nonumber\\
&&t_h=t^{\uparrow,\uparrow}_{3,2}=t^{\downarrow,\downarrow}_{3,2}=
a^*_{2,3}a_{2,2},
\nonumber\\
&&t_f=t^{\uparrow,\uparrow}_{6,5}=t^{\downarrow,\downarrow}_{6,5}=
a^*_{4,6}a_{4,5},
\nonumber\\
&&t_c=t^{\uparrow,\uparrow}_{c}= t^{\uparrow,\uparrow}_{7,4}=a^*_{5,7}a_{5,4}+
{b'}^*_{5,7}b'_{5,4},
\nonumber\\
&&t_c=t^{\downarrow,\downarrow}_{c}= t^{\downarrow,\downarrow}_{7,4}=
a^*_{5,7}a_{5,4}+ b^*_{5,7}b_{5,4}.
\label{Ez4}
\end{eqnarray}
The spin-flip hoppings provide also ten equations as follows
\begin{eqnarray}
&&t^{\uparrow,\downarrow}_{1,5}={b'}^*_{1,1}a_{1,5},
\nonumber\\
&&t^{\downarrow,\uparrow}_{1,5}= \frac{t^{\uparrow,\downarrow}_{1,5}}{\alpha}
={b}^*_{1,1}a_{1,5},
\nonumber\\
&&t^{\uparrow,\downarrow}_{2,1}=a^*_{1,2}b_{1,1},
\nonumber\\
&&t^{\downarrow,\uparrow}_{2,1}=\frac{t^{\uparrow,\downarrow}_{2,1}}{\alpha}=
a^*_{1,2}{b'}_{1,1},
\nonumber\\
&&t^{\uparrow,\downarrow}_{5,4}=a^*_{3,5}b_{3,4},
\nonumber\\
&&t^{\downarrow,\uparrow}_{5,4}=\frac{t^{\uparrow,\downarrow}_{5,4}}{\alpha}=
a^*_{3,5}{b'}_{3,4},
\nonumber\\
&&t^{\uparrow,\downarrow}_{4,3}={b'}^*_{3,4}a_{3,3},
\nonumber\\
&&t^{\downarrow,\uparrow}_{4,3}=\frac{t^{\uparrow,\downarrow}_{4,3}}{\alpha}=
b^*_{3,4}{a}_{3,3},
\nonumber\\
&&t^{\uparrow,\downarrow}_c=t^{\uparrow,\downarrow}_{7,4}={b'}^*_{5,7}a_{5,4}+a^*_{5,7}b_{5,4},
\nonumber\\
&&t^{\downarrow,\uparrow}_c=t^{\downarrow,\uparrow}_{7,4}=\frac{t^{\uparrow,\downarrow}_{7,4}}{
\alpha_c}={b}^*_{5,7}a_{5,4}+a^*_{5,7}{b'}_{5,4}.
\label{Ez5}
\end{eqnarray}
In Eqs(\ref{Ez5}) the coefficients $\alpha,\alpha_c$ take into consideration
the difference between the hoppings $t^{\uparrow,\downarrow}_{n,n'}$ and
$t^{\downarrow,\uparrow}_{n,n'}$ given by the SOI. In the present case, taking into
account only the Rashba term for conducting polymers \cite{Intr12}, one has
$\alpha=\alpha_c=-1$.

The non spin-flip on-site one particle potentials provide eight matching
equations which are as follows:  
\begin{eqnarray}
&&\epsilon_4=\epsilon^{\uparrow,\uparrow}_6=\epsilon^{\downarrow,\downarrow}_6=|a_{4,6}|^2,
\nonumber\\
&&\epsilon_3=\epsilon^{\uparrow,\uparrow}_5=\epsilon^{\downarrow,\downarrow}_5=|a_{4,5}|^2
+|a_{1,5}|^2+|a_{2,5}|^2+|a_{3,5}|^2,
\nonumber\\
&&\epsilon_2=\epsilon^{\uparrow,\uparrow}_2=\epsilon^{\downarrow,\downarrow}_2=|a_{1,2}|^2
+|a_{2,2}|^2,
\nonumber\\
&&\epsilon_2=\epsilon^{\uparrow,\uparrow}_3=\epsilon^{\downarrow,\downarrow}_3=|a_{2,3}|^2
+|a_{3,3}|^2,
\nonumber\\
&&\epsilon_1=\epsilon^{\uparrow,\uparrow}_1=|a_{1,1}|^2
+|a_{5,7}|^2+|b'_{1,1}|^2+|b'_{5,7}|^2,
\nonumber\\
&&\epsilon_1=\epsilon^{\downarrow,\downarrow}_1=|a_{1,1}|^2
+|a_{5,7}|^2+|b_{1,1}|^2+|b_{5,7}|^2,
\nonumber\\
&&\epsilon_1=\epsilon^{\uparrow,\uparrow}_4=|a_{3,4}|^2
+|a_{5,4}|^2+|b'_{3,4}|^2+|b'_{5,4}|^2,
\nonumber\\
&&\epsilon_1=\epsilon^{\downarrow,\downarrow}_4=|a_{3,4}|^2
+|a_{5,4}|^2+|b_{3,4}|^2+|b_{5,4}|^2.
\label{Ez6}
\end{eqnarray}
For spin-flip on-site one particle potential one automatically has
$\epsilon^{\uparrow,\downarrow}_6=\epsilon^{\downarrow,\uparrow}_6=0, \:$
$\epsilon^{\uparrow,\downarrow}_5=\epsilon^{\downarrow,\uparrow}_5=0, \:$
$\epsilon^{\uparrow,\downarrow}_3=\epsilon^{\downarrow,\uparrow}_3=0, \:$
$\epsilon^{\uparrow,\downarrow}_2=\epsilon^{\downarrow,\uparrow}_2=0, \:$
while $\epsilon^{\downarrow,\uparrow}_1=(\epsilon^{\uparrow,\downarrow}_1)^*, \:$
$\epsilon^{\downarrow,\uparrow}_4=(\epsilon^{\uparrow,\downarrow}_4)^*$ is satisfied.
The last two relations further provide two matching equations:
\begin{eqnarray}
&&\epsilon^{\uparrow,\downarrow}_1=a^*_{1,1}b_{1,1}+a^*_{5,7}b_{5,7}+{b'}^*_{1,1}a_{1,1}+
{b'}^*_{5,7}a_{5,7},
\nonumber\\
&&\epsilon^{\uparrow,\downarrow}_4=a^*_{3,4}b_{3,4}+a^*_{5,4}b_{5,4}+{b'}^*_{3,4}a_{3,4}+
{b'}^*_{5,4}a_{5,4}.
\label{Ez7}
\end{eqnarray}
The special treatment of $\epsilon^{\uparrow,\downarrow}_1,\epsilon^{\uparrow,
\downarrow}_4$ is motivated by the fact that the block operators connected to
the bond $4-7$, or sites $1,4$ contain both spin projection components. But
this is done only for the help of the mathematical treatment, and during the
solving process of the matching equations, the left sides of Eqs.(\ref{Ez7})
will be considered zero.

Consequently, one has totally 30 matching equations. These are coupled, complex
algebraic and non-linear equations, which must be solved for the block operator
parameters in order to indeed transform in exact terms the Hamiltonian from
Eq.(\ref{Ez1}), to the Hamiltonian from Eq.(\ref{Ez3}).

Having in mind the transformation to the Hamiltonian form from (\ref{Ez3})
one realizes, that based on the presented procedure another transformation of
the Hamiltonian (\ref{Ez1}) in positive semidefinite form is possible (see
Appendix A), namely
\begin{eqnarray}
\hat H = \sum_{i,z,\sigma} \hat B_{i,z,\sigma} \hat B^{\dagger}_{i,z,\sigma} +
\sum_i \sum_{n=1,2,..6} \hat P_{i,n} + \sum_{<l,k>} V \hat R_{<l,k>} + C,
\label{Ez8}
\end{eqnarray}
where $C=q_U N - N_c \sum_{z,\sigma}y_{z,\sigma} - N_c \sum_{n=1,2,..6} U_n -28 V N_c$,
$N$ being the number of electrons, $N_c$ the number of lattice sites,
$y_{z,\sigma}=\{\hat B_{i,z,\sigma}, \hat B^{\dagger}_{i,z,\sigma}\}$ and $q_U$
being a constant which enters in the matching equations.
In deducing $\hat B_{i,z,\sigma}$, the
procedure described in this section must be applied, the results
being identical, but the Hamiltonian parameters must be changed according to
(\ref{A8}), namely $t^{\sigma,\sigma}_{i,j} \to -t^{\sigma,\sigma}_{i,j}, \quad 
t^{\sigma,\sigma'}_{i,j} \to -t^{\sigma,\sigma'}_{i,j}, \quad
\epsilon^{\sigma,\sigma}_i \to q_U -( \epsilon^{\sigma,\sigma}_i +U_i +2Vr_n), \quad
\epsilon^{\sigma,\sigma'}  \to - \epsilon^{\sigma,\sigma'}$, where $\sigma \ne
\sigma'$ holds. Here, $\hat P_{i,n}=\hat n_{i,n}^{\uparrow}
\hat n_{i,n}^{\downarrow}-(\hat n_{i,n}^{\uparrow}+\hat n_{i,n}^{\downarrow})+1$
represents a positive semidefinite operator which provides its minimum
possible eigenvalue zero, when at least one electron is present at the
site $(i,n)$ (i.e. lattice site i, and in-cell position n). Furthermore,
$\hat R_{<l,k>}= \hat n_l \hat n_k - 2(\hat n_l + \hat n_k) + 4$ is a positive
semidefinite operator which provides its minimum possible eigenvalue zero, when
the $l,k$ nearest neighbor sites are both at least once occupied, but such that
at least one of them is doubly occupied.

At this step I must underline that a given starting Hamiltonian can be
transformed in different positive semidefinite forms, each transformation
placing the corresponding ground state in different regions of the parameter
space \cite{Intr15}.

\subsection{Solution of matching equations in zero external magnetic field.}

The detailed solution of the matching equations is presented in Appendix B.
The final result looks as follows:
\begin{eqnarray}
&&a_{1,1}= \frac{|t|e^{i\theta_1}}{\sqrt{\epsilon_2-t_h}}, \:
a_{1,2}=a_{1,5}=sign(t)\sqrt{\epsilon_2-t_h} \: e^{i\theta_1},
\nonumber\\
&&a_{2,5}=-\frac{\epsilon_2-t_h}{\sqrt{t_h}}e^{i\theta_2}, \:
a_{2,2}=a_{2,3}=\sqrt{t_h} \: e^{i\theta_2}, 
\nonumber\\
&&a_{3,4}=\frac{|t|}{\sqrt{\epsilon_2-t_h}}e^{i\theta_3}, \:
a_{3,3}=a_{3,5}=sign(t)\sqrt{\epsilon_2-t_h} \: e^{i\theta_3},
\nonumber\\
&&a_{4,6}=\sqrt{\epsilon_4} \: e^{i\theta_4}, \: 
a_{4,5}=\frac{t_f}{\sqrt{\epsilon_4}} \: e^{i\theta_4},
\nonumber\\
&&a_{5,4}=\frac{\sqrt{\gamma_1}}{\sqrt{1+\frac{(\sqrt{\gamma_1\gamma_2}
-t_c)^2}{{t^{\uparrow,\downarrow}_c}^2}}} e^ {i\theta_5}, \:
a_{5,7}=\frac{\sqrt{\gamma_2}}{\sqrt{1+\frac{(\sqrt{\gamma_1\gamma_2}-t_c)^2
}{{t^{\uparrow,\downarrow}_c}^2}}} e^{i\theta_5},
\nonumber\\
&&b_{1,1}=sign(t) \frac{t_{2,1}^{\uparrow,\downarrow}}{\sqrt{\epsilon_2-t_h}}e^{i\theta_1},
\: b'_{1,1}=-sign(t) \frac{t_{2,1}^{\uparrow,\downarrow}}{\sqrt{\epsilon_2-t_h}}
e^{i\theta_1},
\nonumber\\
&&b_{3,4}= sign(t) \frac{t_{5,4}^{\uparrow,\downarrow}}{\sqrt{\epsilon_2-t_h}}
e^{i\theta_3}, \: b'_{3,4}= -sign(t) \frac{t_{5,4}^{\uparrow,\downarrow}}{\sqrt{
\epsilon_2-t_h}}e^{i\theta_3},
\nonumber\\
&&b_{5,4}=\frac{\gamma_1\sqrt{\gamma_2}-t_c \sqrt{\gamma_1}}{t^{\uparrow,\downarrow}_c
\sqrt{1+\frac{(\sqrt{\gamma_1\gamma_2}-t_c)^2}{{t^{\uparrow,\downarrow}_c}^2}}}
e^{i\theta_5}, \:
b'_{5,4}=-\frac{\gamma_1\sqrt{\gamma_2}-t_c \sqrt{\gamma_1}}{t^{\uparrow,\downarrow}_c
\sqrt{1+\frac{(\sqrt{\gamma_1\gamma_2}-t_c)^2}{{t^{\uparrow,\downarrow}_c}^2}}}
e^{i\theta_5}, 
\nonumber\\
&&b_{5,7}=\frac{t_c\sqrt{\gamma_2}-\gamma_2 \sqrt{\gamma_1}}{t^{\uparrow,\downarrow}_c
\sqrt{1+\frac{(\sqrt{\gamma_1\gamma_2}-t_c)^2}{{t^{\uparrow,\downarrow}_c}^2}}}
e^{i\theta_5}, \:
b'_{5,7}=-\frac{t_c\sqrt{\gamma_2}-\gamma_2 \sqrt{\gamma_1}}{t^{\uparrow,\downarrow}_c
\sqrt{1+\frac{(\sqrt{\gamma_1\gamma_2}-t_c)^2}{{t^{\uparrow,\downarrow}_c}^2}}}
e^{i\theta_5},
\label{E9}
\end{eqnarray}
where $\theta_1,\theta_2, ...\theta_5$ are arbitrary phases.
The deduced solution is placed in the following parameter space domain
\begin{eqnarray}
&&\epsilon_1 > 0, \: \epsilon_2 > 0, \: \epsilon_3 > 0, \: \epsilon_4 > 0, \:
 t_h > 0, \: 
\epsilon_2-t_h > 0,
\nonumber\\
&&\gamma_1=\epsilon_1 - \frac{t^2+{t^{\uparrow,\downarrow}_{2,1}}^2}{\epsilon_2-t_h} 
> 0, \:
\gamma_2=\epsilon_1 - \frac{t^2+{t^{\uparrow,\downarrow}_{5,4}}^2}{\epsilon_2-t_h} > 0,
\nonumber\\
&&\epsilon_3= \frac{t_f^2}{\epsilon_4} + \frac{\epsilon^2_2-t_h^2}{t_h}, \:
\gamma_1 \gamma_2 =t^2_c +{t^{\uparrow,\downarrow}_c}^2.
\label{E10}
\end{eqnarray}

\subsection{The case of the non-zero external magnetic field}

We analyze also the case when an external magnetic field acts on the system
perpendicular to the plane containing the polymer, i.e, perpendicular to the
unit cell. The effect of the magnetic field is taken into account via i) the
Peierls phase factors attached to the hopping matrix elements in describing
the action on the orbital motion, and ii) the Zeeman term describing the
effect on the spin degrees of freedom. In order
to be explicitly clear, for $B \ne 0$ introducing the notation
$\hat H_{1} = \hat H_K - \hat H_{\epsilon}$, one must transcribe carefully the
$\hat H_{1}$ part of the one particle Hamiltonian as follows
\begin{eqnarray}
\hat H_{1} &=& \sum_{i}\sum_{\sigma,\sigma'} [t_{1,5}^{\sigma,\sigma'} e^{i\phi_{1,5}}
\hat  c^{\dagger}_{i,1,\sigma}\hat c_{i,5,\sigma'} + t_{2,1}^{\sigma,\sigma'}
e^{i\phi_{2,1}} \hat c^{\dagger}_{i,2,\sigma} \hat c_{i,1,\sigma'} + t_{4,3}^{\sigma,\sigma'}
e^{i\phi_{4,3}} \hat c^{\dagger}_{i,4,\sigma} \hat c_{i,3,\sigma'} + t_{5,4}^{\sigma,\sigma'}
e^{i\phi_{5,4}} \hat c^{\dagger}_{i,5,\sigma} \hat c_{i,4,\sigma'}
\nonumber\\
&+& t_{7,4}^{\sigma,\sigma'}
e^{i\phi_{7,4}} \hat c^{\dagger}_{i+a,7,\sigma} \hat c_{i,4,\sigma'} +t_{3,2}^{\sigma,\sigma'}
e^{i\phi_{3,2}} \hat c^{\dagger}_{i,3,\sigma} \hat c_{i,2,\sigma'} + t_{6,5}^{\sigma,\sigma'}
e^{i\phi_{6,5}} \hat c^{\dagger}_{i,6,\sigma} \hat c_{i,5,\sigma'} + h.c. ],
\label{E11}
\end{eqnarray}
hence the total Hamiltonian becomes
\begin{eqnarray}
\hat H = \hat H_1 + \hat H_{\epsilon} + \hat H_U + \hat H_V + \hat H_Z .
\label{E12}
\end{eqnarray}
For the Peierls phase factors one has $\phi_{3,2} = \phi_1, \: \phi_{4,3} =
\phi_{2,1} = \phi_2, \: \phi_{5,4} = \phi_{1,5} = \phi_3, \: \phi_{5,6} =
\phi_{7,4} = 0,$ and $\phi = \phi_1 + 2 \phi_2 + 2 \phi_3 = 2 \pi \Phi/\Phi_0$,
where $\Phi$ is the magnetic flux threading the unit cell, and $\Phi_0=hc/e$
is the flux quantum. Separately for each phase, one has
$\phi_{\alpha}= 2\pi B S_{\alpha}/
\Phi_0$, where $\alpha=1,2,3$, furthermore $S_1 = A(0,2,3), \: S_2 =
A(0,3,4), \: S_3 = A(0,4,5)$, where $A(i,j,k)$ represents the area of the
triangle defined by the points $(i,j,k)$ in Fig.1 presenting the unit cell.
One trivially has $S=S_1+2S_2+2S_3$.

The Zeeman term in Eq.(\ref{E12}) has the standard form
$\hat H_Z = -h \sum_{i,n} (\hat n_{i,n,\uparrow}-\hat n_{i,n,\downarrow})$.
I note that $\hat H_Z$
maintains the structure of $\hat H$ since only it renormalizes the on-site
one particle potentials in $\hat H_{\epsilon}$
by $\epsilon^{\sigma,\sigma}_i \to \epsilon^{\sigma,\sigma}_i
+ \mu  h$, where $\mu = -1$ for $\sigma = \uparrow$ and $\mu = +1$  for
$\sigma = \downarrow$.

\subsection{Matching equations in non-zero external magnetic field.}

In order to transform the Hamiltonian from (\ref{E12}) in positive semidefinite
form one uses the same block operators as before, given in (\ref{Ez2}).
The transformation in positive semidefinite form of the $\hat H$ can be given
with the condition to consider in Eq.(\ref{Ez2}) different $a_{k,n}$
coefficients for $\hat B_{i,k,\uparrow}$ (these will be denoted by $a_{k,n,u}$) and
$\hat B_{i,k,\downarrow}$ (these will be denoted by $a_{k,n,d} \ne a_{k,n,u}$), e.g.
\begin{eqnarray}
&&\hat B_{i,1,\uparrow}= a_{1,1,u} \hat c_{i,1,\uparrow} +a_{1,2,u} \hat c_{i,2,\uparrow} +
a_{1,5,u} \hat c_{i,5,\uparrow} + b_{1,1} \hat c_{i,1,\downarrow},
\nonumber\\
&&\hat B_{i,1,\downarrow}= a_{1,1,d} \hat c_{i,1,\downarrow} +a_{1,2,d}
\hat c_{i,2,\downarrow} + a_{1,5,d} \hat c_{i,5,\downarrow} + b'_{1,1}
\hat c_{i,1,\uparrow}, \: etc.
\label{E13}
\end{eqnarray}
The transformation in the form positive semidefinite form given in (\ref{Ez3}),
or (\ref{Ez8}) (with the conditions mentioned below (\ref{Ez8})) requires the
following matching equations:

The first ten matching equations from (\ref{Ez4}) for hoppings without spin-flip
become now 18 equations (18, since also in Eq.(\ref{Ez4}) $t_c^{\uparrow,\uparrow}$
and $t_c^{\downarrow,\downarrow}$ had different equations) as follows
\begin{eqnarray}
&&t^{\uparrow,\uparrow}_{1,5} e^{i\phi_{1,5}} = te^{i\phi_3} =
a^*_{1,1,u}a_{1,5,u},
\nonumber\\
&&t^{\downarrow,\downarrow}_{1,5} e^{i\phi_{1,5}} = te^{i\phi_3} =
a^*_{1,1,d}a_{1,5,d},
\nonumber\\
&&t^{\uparrow,\uparrow}_{2,1} e^{i\phi_{2,1}}   = te^{i\phi_2} =
a^*_{1,2,u}a_{1,1,u},
\nonumber\\
&&t^{\downarrow,\downarrow}_{2,1} e^{i\phi_{2,1}}   = te^{i\phi_2} =
a^*_{1,2,d}a_{1,1,d},
\nonumber\\
&&t^{\uparrow,\uparrow}_{4,3} e^{i\phi_{4,3}}    = te^{i\phi_2} =
a^*_{3,4,u}a_{3,3,u},
\nonumber\\
&&t^{\downarrow,\downarrow}_{4,3} e^{i\phi_{4,3}}    = te^{i\phi_2} =
a^*_{3,4,d}a_{3,3,d},
\nonumber\\
&&t^{\uparrow,\uparrow}_{5,4} e^{i\phi_{5,4}}    = te^{i\phi_3} =
a^*_{3,5,u}a_{3,4,u},
\nonumber\\
&&t^{\downarrow,\downarrow}_{5,4} e^{i\phi_{5,4}}    = te^{i\phi_3} =
a^*_{3,5,d}a_{3,4,d},
\nonumber\\
&&t^{\uparrow,\uparrow}_{3,2}  e^{i\phi_{3,2}}   = t_he^{i\phi_1} =
a^*_{2,3,u}a_{2,2,u},
\nonumber\\
&&t^{\downarrow,\downarrow}_{3,2}  e^{i\phi_{3,2}}   = t_he^{i\phi_1} =
a^*_{2,3,d}a_{2,2,d},
\nonumber\\
&&t^{\uparrow,\uparrow}_{6,5} e^{i\phi_{6,5}}  = t_f =
a^*_{4,6,u}a_{4,5,u},
\nonumber\\
&&t^{\downarrow,\downarrow}_{6,5} e^{i\phi_{6,5}}  = t_f =
a^*_{4,6,d}a_{4,5,d},
\nonumber\\
&&t^{\uparrow,\uparrow}_{7,4}  e^{i\phi_{7,4}} = t_c =a^*_{5,7,u}a_{5,4,u}+
{b'}^*_{5,7}b'_{5,4},
\nonumber\\
&&t^{\downarrow,\downarrow}_{7,4} e^{i\phi_{7,4}}  =t_c =
a^*_{5,7,d}a_{5,4,d}+ b^*_{5,7}b_{5,4},
\nonumber\\
&&0=t^{\uparrow,\uparrow}_{2,5}=
a^*_{1,2,u}a_{1,5,u}+a^*_{2,2,u}a_{2,5,u},
\nonumber\\
&&0=t^{\downarrow,\downarrow}_{2,5}=
a^*_{1,2,d}a_{1,5,d}+a^*_{2,2,d}a_{2,5,d},
\nonumber\\
&&0=t^{\uparrow,\uparrow}_{3,5}=
a^*_{3,3,u}a_{3,5,u}+a^*_{2,3,u}a_{2,5,u},
\nonumber\\
&&0=t^{\downarrow,\downarrow}_{3,5}=
a^*_{3,3,d}a_{3,5,d}+a^*_{2,3,d}a_{2,5,d}.
\label{E14}
\end{eqnarray}
The second group of ten matching equations relating the spin-flip contributions
describing the SOI interaction, and substituting (\ref{Ez5}), become
\begin{eqnarray}
&&t^{\uparrow,\downarrow}_{1,5}e^{i\phi_{1,5}} = -\lambda e^{i\phi_3} =
{b'}^*_{1,1}a_{1,5,d},
\nonumber\\
&&t^{\downarrow,\uparrow}_{1,5}e^{i\phi_{1,5}} = \frac{t^{\uparrow,\downarrow}_{1,5}
e^{i\phi_{1,5}} }{\alpha} = \lambda  e^{i\phi_3}=
={b}^*_{1,1}a_{1,5,u},
\nonumber\\
&&t^{\uparrow,\downarrow}_{2,1} e^{i\phi_{2,1}}  =\lambda e^{i\phi_2}=
a^*_{1,2,u}b_{1,1},
\nonumber\\
&&t^{\downarrow,\uparrow}_{2,1}e^{i\phi_{2,1}}  =\frac{t^{\uparrow,\downarrow}_{2,1}
e^{i\phi_{2,1}} }{\alpha}=- \lambda  e^{i\phi_2} =
a^*_{1,2,d}{b'}_{1,1},
\nonumber\\
&&t^{\uparrow,\downarrow}_{5,4}e^{i\phi_{5,4}}  =-\lambda e^{i\phi_3}=
a^*_{3,5,u}b_{3,4},
\nonumber\\
&&t^{\downarrow,\uparrow}_{5,4}e^{i\phi_{5,4}}  =\frac{t^{\uparrow,\downarrow}_{5,4}
e^{i\phi_{5,4}}  }{\alpha}= \lambda e^{i\phi_3} =
a^*_{3,5,d}{b'}_{3,4},
\nonumber\\
&&t^{\uparrow,\downarrow}_{4,3}e^{i\phi_{4,3}}  =\lambda e^{i\phi_2}=
{b'}^*_{3,4}a_{3,3,d},
\nonumber\\
&&t^{\downarrow,\uparrow}_{4,3}e^{i\phi_{4,3}}  =\frac{t^{\uparrow,\downarrow}_{4,3}
e^{i\phi_{4,3}}  }{\alpha}= - \lambda e^{i\phi_2} =
b^*_{3,4}{a}_{3,3,u},
\nonumber\\
&&t^{\uparrow,\downarrow}_{7,4} e^{i\phi_{4,7}} =\lambda_c=
{b'}^*_{5,7}a_{5,4,d}+a^*_{5,7,u}b_{5,4},
\nonumber\\
&&t^{\downarrow,\uparrow}_{7,4}e^{i\phi_{7,4}}  =\frac{t^{\uparrow,\downarrow}_{7,4}
  e^{i\phi_{7,4}}  }{\alpha_c} = - \lambda_c =
    {b}^*_{5,7}a_{5,4,u}+a^*_{5,7,d}{b'}_{5,4}.
\label{E15}
\end{eqnarray}
Here, taking into account symmetry considerations, we denoted the spin-flip
hopping strengths by $\lambda$ and $\lambda_c$ according to the relations
$t^{\uparrow,\downarrow}_{1,5}e^{i\phi_{1,5}}  =-\lambda e^{i\phi_3}, \:
t^{\uparrow,\downarrow}_{5,4}e^{i\phi_{5,4}}  =-\lambda e^{i\phi_3}, \:
t^{\uparrow,\downarrow}_{2,1}e^{i\phi_{2,1}}  =\lambda e^{i\phi_2}, \:
t^{\uparrow,\downarrow}_{4,3}e^{i\phi_{4,3}}  =\lambda e^{i\phi_2}, \:
t^{\uparrow,\downarrow}_{7,4}e^{i\phi_{7,4}}  =\lambda_c e^{i\phi_{7,4}}=\lambda_c.$
Finally, the last ten matching equations provided by the on-site one-particle
potentials are now 14 equations, as follows
\begin{eqnarray}
&&\epsilon^{\uparrow,\uparrow}_6 = \epsilon_6 - h = |a_{4,6,u}|^2,
\nonumber\\
&&\epsilon^{\downarrow,\downarrow}_6 = \epsilon_6 + h = |a_{4,6,d}|^2,
\nonumber\\
&&\epsilon^{\uparrow,\uparrow}_5= \epsilon_5 -h = |a_{4,5,u}|^2
+|a_{1,5,u}|^2+|a_{2,5,u}|^2+|a_{3,5,u}|^2,
\nonumber\\
&&\epsilon^{\downarrow,\downarrow}_5= \epsilon_5 + h =
|a_{4,5,d}|^2 +|a_{1,5,d}|^2+|a_{2,5,d}|^2+|a_{3,5,d}|^2,
\nonumber\\
&&\epsilon^{\uparrow,\uparrow}_2= \epsilon_2 - h = |a_{1,2,u}|^2 + |a_{2,2,u}|^2,
\nonumber\\
&&\epsilon^{\downarrow,\downarrow}_2= \epsilon_2 + h =
|a_{1,2,d}|^2 + |a_{2,2,d}|^2,
\nonumber\\
&&\epsilon^{\uparrow,\uparrow}_3= \epsilon_3 -h =
|a_{2,3,u}|^2+|a_{3,3,u}|^2,
\nonumber\\
&&\epsilon^{\downarrow,\downarrow}_3= \epsilon_3 + h =
|a_{2,3,d}|^2+|a_{3,3,d}|^2,
\nonumber\\
&&\epsilon^{\uparrow,\uparrow}_1 = \epsilon_1 -h =
|a_{1,1,u}|^2 + |a_{5,7,u}|^2+|b'_{1,1}|^2 +|b'_{5,7}|^2,
\nonumber\\
&&\epsilon^{\downarrow,\downarrow}_1 = \epsilon_1 + h =
|a_{1,1,d}|^2 + |a_{5,7,d}|^2+|b_{1,1}|^2+|b_{5,7}|^2,
\nonumber\\
&&\epsilon^{\uparrow,\uparrow}_4 = \epsilon_4 - h =
|a_{3,4,u}|^2 + |a_{5,4,u}|^2+|b'_{3,4}|^2+|b'_{5,4}|^2,
\nonumber\\
&&\epsilon^{\downarrow,\downarrow}_4 = \epsilon_4 + h =
|a_{3,4,d}|^2 + |a_{5,4,d}|^2+|b_{3,4}|^2+|b_{5,4}|^2,
\nonumber\\
&&\epsilon^{\uparrow,\downarrow}_1 = 0 = a^*_{1,1,u}b_{1,1} + a^*_{5,7,u} b_{5,7} +
{b'}^*_{1,1}a_{1,1,d} + {b'}^*_{5,7}a_{5,7,d},
\nonumber\\
&&\epsilon^{\uparrow,\downarrow}_4 = 0 = a^*_{3,4,u}b_{3,4} + a^*_{5,4,u}b_{5,4} +
{b'}^*_{3,4}a_{3,4,d} + {b'}^*_{5,4}a_{5,4,d},
\label{E16}
\end{eqnarray}
where one has only 14 equations, since in Eqs.(\ref{Ez6},\ref{Ez7}), the two
equalities in (\ref{Ez7}) and the last four equalities from (\ref{Ez6})
(so totally six equalities) provide only six equations in (\ref{E16}), i.e.
not multiplies.

Finally, one has in the ${\bf B} \ne 0$ case 42 coupled, non-linear complex
algebraic matching equations. I only mention, that numerical software for
solving such system of equations is not present today. 

\subsection{Solution of the matching equations in the presence of the external
 magnetic field}

The matching equations at ${\bf B} \ne 0$ are solved in the Appendix C. The
solution looks as follows:

In the case of the first eight block operators, the block operator parameters
are given by the following relations:
The coefficients of the $z=1$ block operators $\hat B_{i,z=1,\sigma}$ are given by
\begin{eqnarray}
&&a_{1,1,u}= [ \frac{ t^2 (T_h + \epsilon_3 -h)}{(\epsilon_2-h)(\epsilon_3-h) -
t_h^2} ]^{1/2} e^{i \chi_{1,u}}, \quad \quad a_{1,1,d}= [ \frac{ t^2 (T_h +
\epsilon_3 +h)}{(\epsilon_2+h)(\epsilon_3+h) - t_h^2} ]^{1/2} e^{i \chi_{1,d}},
\nonumber\\
&&a_{1,5,u} = t e^{i \phi_3} [ \frac{(\epsilon_2-h)(\epsilon_3-h) -t_h^2}{
t^2 (T_h + \epsilon_3 -h)} ]^{1/2} e^{i \chi_{1,u}}, \quad a_{1,5,d}=
t e^{i \phi_3} [ \frac{(\epsilon_2+h)(\epsilon_3+h) -t_h^2}{
t^2 (T_h + \epsilon_3 +h)} ]^{1/2} e^{i \chi_{1,d}},
\nonumber\\
&&a_{1,2,u}=  t e^{-i \phi_2} [ \frac{(\epsilon_2-h)(\epsilon_3-h) -t_h^2}{
t^2 (T_h + \epsilon_3 -h)} ]^{1/2} e^{i \chi_{1,u}}, \quad
a_{1,2,d}=  t e^{-i \phi_2} [ \frac{(\epsilon_2+h)(\epsilon_3+h) -t_h^2}{
t^2 (T_h + \epsilon_3 +h)} ]^{1/2} e^{i \chi_{1,d}}, 
\nonumber\\
&&b_{1,1} = \frac{\lambda}{t} \: [ \frac{ t^2 (T_h + \epsilon_3 -h)}{(\epsilon_2-
h)(\epsilon_3-h) -t_h^2} ]^{1/2} e^{i \chi_{1,u}}, \quad
b'_{1,1} = - \frac{\lambda}{t} \: [ \frac{ t^2 (T_h + \epsilon_3 +h)}{(\epsilon_2+
h)(\epsilon_3+h) -t_h^2} ]^{1/2} e^{i \chi_{1,d}},
\label{E17}
\end{eqnarray}
where $T_h=t_h e^{-i\phi} \geq 0$ must be satisfied. 
The coefficients of the $z=2$ block operators become
\begin{eqnarray}
&&a_{2,2,u}= [\frac{(\epsilon_3-h)T_h + t_h^2}{T_h+\epsilon_3-h}]^{1/2}
e^{i \chi_{2,u}}, \quad \quad
\quad a_{2,2,d}= [\frac{(\epsilon_3+h)T_h + t_h^2}{T_h+\epsilon_3+h}]^{1/2}
e^{i \chi_{2,d}},
\nonumber\\   
&&a_{2,3,u}=t_h e^{-i\phi_1} [ \frac{T_h+\epsilon_3-h}{(\epsilon_3-h)T_h + t_h^2}
]^{1/2} e^{i \chi_{2,u}}, \quad a_{2,3,d}=t_h e^{-i\phi_1} [ \frac{T_h+\epsilon_3+h}{
(\epsilon_3+h)T_h + t_h^2} ]^{1/2} e^{i \chi_{2,d}}, 
\nonumber\\
&&a_{2,5,u}= -e^{i(\phi_2+\phi_3)} \frac{[(\epsilon_2-h)(\epsilon_3-h)-t_h^2]}{
  \sqrt{T_h(\epsilon_2-h)+t_h^2}\sqrt{\epsilon_3-h +T_h}} \:  e^{i \chi_{2,u}},
\nonumber\\
&&a_{2,5,d}= -e^{i(\phi_2+\phi_3)} \frac{[(\epsilon_2+h)(\epsilon_3+h)-t_h^2]}{
\sqrt{T_h(\epsilon_2+h)+t_h^2}\sqrt{\epsilon_3+h +T_h}} \:  e^{i \chi_{2,d}}. 
\label{E18}
\end{eqnarray}
The coefficients of the $z=3$ block operators become
\begin{eqnarray}
&&a_{3,3,u}=te^{i\phi_2} \big[\frac{T_h[(\epsilon_2-h)(\epsilon_3-h)-t_h^2]}{
t^2[T_h(\epsilon_2-h)+t_h^2]}\big]^{1/2} e^{i \chi_{3,u}}, \quad
a_{3,3,d}=te^{i\phi_2} \big[\frac{T_h[(\epsilon_2+h)(\epsilon_3+h)-t_h^2]}{
t^2[T_h(\epsilon_2+h)+t_h^2]}\big]^{1/2} e^{i \chi_{3,d}},
\nonumber\\  
&&a_{3,4,u}=\big[\frac{t^2[T_h(\epsilon_2-h)+t_h^2]}{T_h[(\epsilon_2-h)(\epsilon_3
-h)-t_h^2]} \big]^{1/2} e^{i \chi_{3,u}}, \quad
a_{3,4,d}=\big[\frac{t^2[T_h(\epsilon_2+h)+t_h^2]}{T_h[(\epsilon_2+h)(\epsilon_3
+h)-t_h^2]} \big]^{1/2} e^{i \chi_{3,d}}, 
\nonumber\\
&&a_{3,5,u} = t e^{-i\phi_3}\big[\frac{T_h[(\epsilon_2-h)(\epsilon_3-h)-t_h^2]}{
t^2[T_h(\epsilon_2-h)+t_h^2]}\big]^{1/2} e^{i \chi_{3,u}}, \quad
a_{3,5,d} = t e^{-i\phi_3}\big[\frac{T_h[(\epsilon_2+h)(\epsilon_3+h)-t_h^2]}{
t^2[T_h(\epsilon_2+h)+t_h^2]}\big]^{1/2} e^{i \chi_{3,d}},
\nonumber\\
&&b_{3,4}= -\frac{\lambda}{t} \big[\frac{t^2[T_h(\epsilon_2-h)+t_h^2]}{
T_h[(\epsilon_2-h)(\epsilon_3-h)-t_h^2]} \big]^{1/2} e^{i \chi_{3,u}}, \quad 
b'_{3,4} = \frac{\lambda}{t}\big[\frac{t^2[T_h(\epsilon_2+h)+t_h^2]}{
T_h[(\epsilon_2+h)(\epsilon_3+h)-t_h^2]} \big]^{1/2} e^{i \chi_{3,d}}.
\label{E19}
\end{eqnarray}
The coefficients of the $z=4$ block operators are the following ones
\begin{eqnarray}
&&a_{4,5,u}=\frac{t_f}{\sqrt{\epsilon_6-h}}  e^{i \chi_{4,u}}, \quad
a_{4,5,d}=\frac{t_f}{\sqrt{\epsilon_6+h}}  e^{i \chi_{4,d}},
\nonumber\\
&&a_{4,6,u}= \sqrt{\epsilon_6-h} e^{i \chi_{4,u}}, \quad
a_{4,6,d}= \sqrt{\epsilon_6+h} e^{i \chi_{4,d}}. 
\label{E20}
\end{eqnarray}
The calculation of the block operator coefficients connected to the block
operators $\hat B_{i,5,\sigma}$, i.e. $z=5$ is much more tricky. First we should
introduce the following notations
\begin{eqnarray}
&& I_{1,u} =\epsilon_1-h -|a_{1,1,u}|^2-|b'_{1,1}|^2, \quad
I_{1,d} =\epsilon_1+h -|a_{1,1,d}|^2-|b_{1,1}|^2,
\nonumber\\
&& I_{2,u} =\epsilon_4-h -|a_{3,4,u}|^2-|b'_{3,4}|^2, \quad
I_{2,d} =\epsilon_4+h -|a_{3,4,d}|^2-|b_{3,4}|^2,
\nonumber\\
&& V_1 = -a^*_{1,1,u}b_{1,1} -{b'}^*_{1,1} a_{1,1,d}, \quad
V_2 =  -a^*_{3,4,u}b_{3,4} -{b'}^*_{3,4} a_{3,4,d}.
\label{E21}
\end{eqnarray}
Given by (\ref{E17}-\ref{E20}), all quantities from (\ref{E21}) have known
and real values. Now, based on (\ref{E21}) one defines the expressions
\begin{eqnarray}
&&W_1=\frac{t_c-z I_{1,u}}{V_2-z \lambda_c}, \quad
W_2 =\frac{z t_c -I_{2,u}}{z \lambda_c - V_2},
\nonumber\\
&&W_3=\frac{t_c- y I_{1,d}}{V_2 + y \lambda_c}, \quad
W_4 = \frac{I_{2,d}- y t_c}{V_2 + y \lambda_c}, \quad
y=\frac{z \lambda_c -V_2}{\lambda_c + z V_1}.
\label{E22}
\end{eqnarray}
Based on (\ref{E22}) the following relations providing the requested remaining
block operator parameters hold
\begin{eqnarray}
&&b'_{5,7} = W_1 a_{5,4,d}, \quad b'_{5,4} = W_2 a_{5,4,d}, \quad
b^*_{5,7} = W_3 a^*_{5,4,u}, \quad b^*_{5,4} = W_4 a^*_{5,4,u},
\nonumber\\
&&a^*_{5,7,u} = \frac{a^*_{5,4,u}}{z}, \quad a_{5,7,d} = \frac{a_{5,4,d}}{y}
\label{E23}
\end{eqnarray}
The remaining two block operator parameters are given by
\begin{eqnarray}
a_{5,4,u} = \big[\frac{V_2 W_1 -y V_1 W_2}{W_4 W_1 -W_3 W_2(y/z)} \big]^{1/2}
e^{i \chi_{5,u}}, \quad
a_{5,4,d} = \big[\frac{V_2 W_3 -z V_1 W_4}{W_2 W_3 -W_1 W_4(z/y)} \big]^{1/2}
e^{i \chi_{5,d}},
\label{E24}
\end{eqnarray}
In the provided expressions Eqs.(\ref{E17}-\ref{E24}), the phases
$\chi_{k,u},\chi_{k,d}$, $k=1,2,...5$ are arbitrary, and z is an arbitrary
real parameter restricted by the conditions $|a_{5,4,u}|^2, |a_{5,4,d}|^2  > 0$.

The parameter space region where the solution is valid is fixed by the relations
$I_{\alpha,\nu} \geq 0$, where $\alpha=1,2$, $\nu=u,d$. These relations provide
lower limits for the on-site one particle potentials $\epsilon_1$ and
$\epsilon_4$. Furthermore on has $T_h \geq 0$, and two remaining matching
equations from (\ref{E16}), namely
$\epsilon_5 + \mu(\nu) h = |a_{4,5,\nu}|^2 + |a_{1,5,\nu}|^2 + |a_{2,5,\nu}|^2 +
|a_{3,5,\nu}|^2$, where $\mu(u)=-1$ and $\mu(d)=+1$ which relates the $\epsilon_5$
value.

\section{Exact ground states}

Once we solved the matching equations, the exact transformation of the
Hamiltonian in positive semidefinite form has been performed. Now based on the
positive semidefinite form of the Hamiltonian, one can deduce the exact ground
states. This step will be done on two lines. First we deduce the exact ground
state corresponding to the transformed Hamiltonian (\ref{Ez3}), and in the
second step, the exact ground state corresponding to the transformed Hamiltonian
from (\ref{Ez8}).

\begin{figure}[h]
\includegraphics[height=5cm, width=11cm]{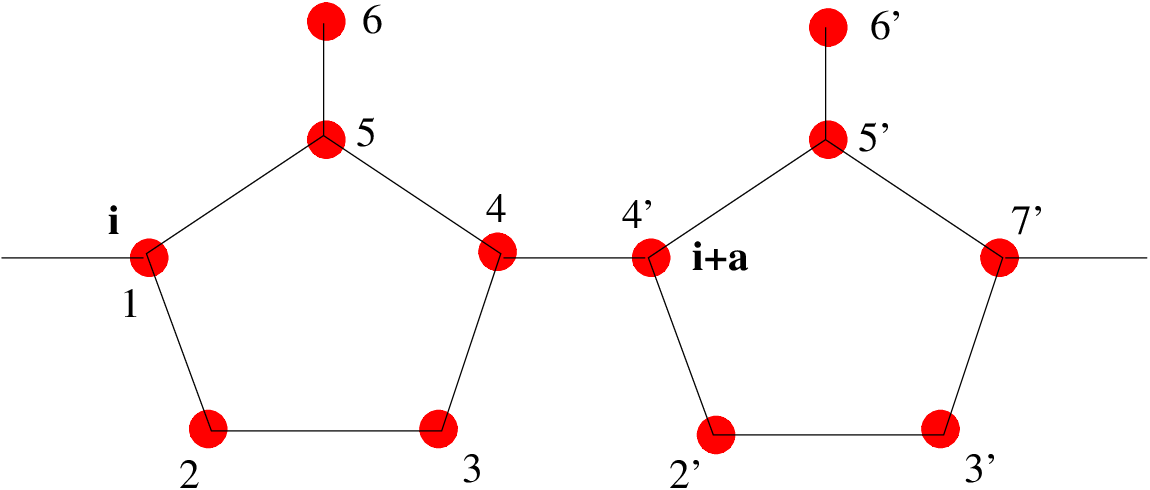}
\caption{The $\hat X^{\dagger}_i$ operator connected to the lattice site ${\bf i}$.
The red dotes denote the sites that are belong to  $\hat X^{\dagger}_i$. }
\end{figure}

\subsection{Ground states in the low density case}

One concentrates now on the transformed Hamiltonian $\hat H$ from (\ref{Ez3})
used with periodic boundary conditions.
The procedure is based on the deduction of a local operator $\hat X_i^{\dagger}$
that satisfy the property
\begin{eqnarray}
\{ \hat B_{j,z,\sigma}, \hat X_i^{\dagger} \} =0
\label{E25}
\end{eqnarray}  
for all $j,i$ lattice sites, all $z=1,2,..5$, and all $\sigma$. The logic
in this choice is that based on
(\ref{E25}), one has for $|\Psi\rangle = \prod_i \hat X_i^{\dagger}|0\rangle$,
where $|0\rangle$ is the vacuum state with no electrons present, the property
\begin{eqnarray}  
(\sum_{i,z,\sigma} \hat B^{\dagger}_{i,z,\sigma} \hat B_{i,z,\sigma}) |\Psi\rangle =0
\label{E26}
\end{eqnarray}
holds (since all $\hat B_{i,z,\sigma}$ can be interchanged with
$\hat X_i^{\dagger}$). Consequently, introducing $|\Psi\rangle$ in the kernel of
$\hat H_U$ (i.e. the Hilbert subspace provided by all $|\chi\rangle$ for which
$\hat H_U |\chi\rangle =0$), and in the kernel of $\hat H_V$,
one has the ground state.

In deducing $\hat X_i^{\dagger}$, one takes into consideration a most general
domain defined on two unit cells, as presented in Fig.2. The operator is
defined as
\begin{eqnarray}
  \hat X^{\dagger}_i &=&x^*_{1,\uparrow} \hat c^{\dagger}_{i,1,\uparrow} +
x^*_{2,\uparrow} \hat c^{\dagger}_{i,2,\uparrow} +  
x^*_{3,\uparrow} \hat c^{\dagger}_{i,3,\uparrow} +
x^*_{4,\uparrow} \hat c^{\dagger}_{i,4,\uparrow} +
x^*_{5,\uparrow} \hat c^{\dagger}_{i,5,\uparrow} +
x^*_{6,\uparrow} \hat c^{\dagger}_{6,1,\uparrow} +
x^*_{4',\uparrow} \hat c^{\dagger}_{i+1,1,\uparrow}
\nonumber\\
&+&
x^*_{2',\uparrow} \hat c^{\dagger}_{i+1,2,\uparrow} +
x^*_{3',\uparrow} \hat c^{\dagger}_{i+1,3,\uparrow} +
x^*_{5',\uparrow} \hat c^{\dagger}_{i+1,5,\uparrow} +
x^*_{6',\uparrow} \hat c^{\dagger}_{i+1,6,\uparrow} +
x^*_{7',\uparrow} \hat c^{\dagger}_{i+1,4,\uparrow} +
y^*_{1,\downarrow} \hat c^{\dagger}_{i,1,\downarrow}
\nonumber\\
&+&
y^*_{2,\downarrow} \hat c^{\dagger}_{i,2,\downarrow} +
y^*_{3,\downarrow} \hat c^{\dagger}_{i,3,\downarrow} +
y^*_{4,\downarrow} \hat c^{\dagger}_{i,4,\downarrow} +
y^*_{5,\downarrow} \hat c^{\dagger}_{i,5,\downarrow} +
y^*_{6,\downarrow} \hat c^{\dagger}_{i,6,\downarrow} +
y^*_{4',\downarrow} \hat c^{\dagger}_{i+1,1,\downarrow} +
y^*_{2',\downarrow} \hat c^{\dagger}_{i+1,2,\downarrow}
\nonumber\\
&+&
y^*_{3',\downarrow} \hat c^{\dagger}_{i+1,3,\downarrow} +
y^*_{5',\downarrow} \hat c^{\dagger}_{i+1,5,\downarrow} +
y^*_{6',\downarrow} \hat c^{\dagger}_{i+1,6,\downarrow} +
y^*_{7',\downarrow} \hat c^{\dagger}_{i+1,4,\downarrow}.
\label{E27}
\end{eqnarray}
Now the (\ref{E25}) anticommutators are presented, calculated in order with
$\hat B_{i,1,\uparrow}, \hat B_{i,1,\downarrow}, \hat B_{i,2,\uparrow},
\hat B_{i,2,\downarrow}, ..., \hat B_{i,5,\uparrow},\hat B_{i,5,\downarrow},  
\hat B_{i+1,1,\uparrow},\hat B_{i+1,1,\downarrow}, \hat B_{i+1,2,\uparrow},
\hat B_{i+1,2,\downarrow}, ..., \hat B_{i+1,5,\uparrow}, \hat B_{i+1,5,\downarrow}$,   
$\hat B_{i-1,5,\uparrow},\hat B_{i-1,5,\downarrow}$, in total 22 equations
\begin{eqnarray}
&&x^*_{2,\uparrow} a_{1,2,u} + x^*_{5,\uparrow} a_{1,5,u} + x^*_{1,\uparrow} a_{1,1,u} +  
y^*_{1,\downarrow} b_{1,1} = 0, 
\nonumber\\
&&y^*_{1,\downarrow} a_{1,1,d} + y^*_{5,\downarrow} a_{1,5,d} + y^*_{2,\downarrow} a_{1,2,d}
+  x^*_{1,\uparrow} b'_{1,1} = 0,
\nonumber\\
&&x^*_{2,\uparrow} a_{2,2,u} + x^*_{3,\uparrow} a_{2,3,u} + x^*_{5,\uparrow} a_{2,5,u} = 0, 
\nonumber\\
&&y^*_{3,\downarrow} a_{2,3,d} + y^*_{5,\downarrow} a_{2,5,d} + y^*_{2,\downarrow} a_{2,2,d}
= 0,
\nonumber\\
&&x^*_{3,\uparrow} a_{3,3,u} + x^*_{4,\uparrow} a_{3,4,u} + x^*_{5,\uparrow} a_{3,5,u} +  
y^*_{4,\downarrow} b_{3,4} = 0, 
\nonumber\\
&&y^*_{3,\downarrow} a_{3,3,d} + y^*_{4,\downarrow} a_{3,4,d} + y^*_{5,\downarrow} a_{3,5,d}
+  x^*_{4,\uparrow} b'_{3,4} = 0,
\nonumber\\
&&x^*_{5,\uparrow} a_{4,5,u} + x^*_{6,\uparrow} a_{4,6,u} = 0, 
\nonumber\\
&&y^*_{5,\downarrow} a_{4,5,d} + y^*_{6,\downarrow} a_{4,6,d} = 0,
\nonumber\\
&&x^*_{4,\uparrow} a_{5,4,u} + x^*_{4',\uparrow} a_{5,7,u} + y^*_{4,\downarrow} b_{5,4} +  
y^*_{4',\downarrow} b_{5,7} = 0, 
\nonumber\\
&&x^*_{4,\uparrow} b'_{5,4} + x^*_{4',\uparrow} b'_{5,7} + y^*_{4,\downarrow} a_{5,4,d}
+  y^*_{4',\downarrow} a_{5,7,d} = 0,
\nonumber\\
&&x^*_{4',\uparrow} a_{1,1,u} + x^*_{2',\uparrow} a_{1,2,u} + x^*_{5',\uparrow} a_{1,5,u} +  
y^*_{4',\downarrow} b_{1,1} = 0, 
\nonumber\\
&&x^*_{4',\uparrow} b'_{1,1} + y^*_{4',\downarrow} a_{1,1,d} + y^*_{2',\downarrow} a_{1,2,d}
+  y^*_{5',\downarrow} a_{1,5,d} = 0,
\nonumber\\
&&x^*_{2',\uparrow} a_{2,2,u} + x^*_{3',\uparrow} a_{2,3,u} + x^*_{5',\uparrow} a_{2,5,u} =  
0, 
\nonumber\\
&&y^*_{2',\downarrow} a_{2,2,d} + y^*_{3',\downarrow} a_{2,3,d} + y^*_{5',\downarrow}
a_{2,5,d} = 0,
\nonumber\\
&&x^*_{3',\uparrow} a_{3,3,u} + x^*_{5',\uparrow} a_{3,5,u} + x^*_{7',\uparrow}
a_{3,4,u} +  y^*_{7',\downarrow} b_{3,4} = 0, 
\nonumber\\
&&y^*_{3',\downarrow} a_{3,3,d} + y^*_{5',\downarrow} a_{3,5,d} + x^*_{7',\uparrow} b'_{3,4}
+  y^*_{7',\downarrow} a_{3,4,d} = 0,
\nonumber\\
&&x^*_{5',\uparrow} a_{4,5,u} + x^*_{6',\uparrow} a_{4,6,u} = 0, 
\nonumber\\
&&y^*_{5',\downarrow} a_{4,5,d} +  y^*_{6',\downarrow} a_{4,6,d} = 0,
\nonumber\\
&&x^*_{7',\uparrow} a_{5,4,u} + y^*_{7',\downarrow} b_{5,4} = 0, 
\nonumber\\
&&x^*_{7',\uparrow} b'_{5,4} +  y^*_{7',\downarrow} a_{5,4,d} = 0,
\nonumber\\
&&x^*_{1,\uparrow} a_{5,7,u} + y^*_{1,\downarrow} b_{5,7} = 0, 
\nonumber\\
&&x^*_{1,\uparrow} b'_{5,7} +  y^*_{1,\downarrow} a_{5,7,d} = 0,
\label{E28}  
\end{eqnarray}
In the ${\bf B}=0$ case when $a_{i,j,u}=a_{i,j,d}=a_{i,j}$, $h=\phi=0$ holds,
because of
$b'_{5,4}=-b_{5,4}$ and $b'_{5,7}=-b_{5,7}$,  (see (\ref{E9})), the last four
equations from (\ref{E28}) provide $x_{1,\uparrow}=y_{1,\downarrow}=x_{7',\uparrow}
=y_{7',\downarrow}=0$, which is not satisfied automatically at ${\bf B} \ne 0$,
i.e. $a_{i,j,u} \ne  a_{i,j,d}$. That is why, (\ref{E28}) must be separately
solved in the presence, and in the absence of the external magnetic field.

\subsubsection{Solution for $\hat X^{\dagger}_i$ at $\hat B =0$}

The solutions of (\ref{E28}) deduced in Appendix D, can be given as follows
\begin{eqnarray} 
&&x^*_{6,\uparrow} = \frac{t_f}{\epsilon_4} x^*_{2,\uparrow}, \:
x^*_{6',\uparrow} =  \frac{t_f}{\epsilon_4} x^*_{3',\uparrow}, \:
x^*_{5,\uparrow} = - x^*_{2,\uparrow}, \: x^*_{5',\uparrow} = - x^*_{3',\uparrow}, \: 
x^*_{3,\uparrow} = \frac{-\epsilon_2}{t_h} x^*_{2,\uparrow}, \: x^*_{1,\uparrow}=0,
\nonumber\\
&&x^*_{2',\uparrow} = \frac{-\epsilon_2}{t_h} x^*_{3',\uparrow}, \:
x^*_{4,\uparrow} = \frac{\epsilon_2^2 - t^2_h}{t_h(t^2+\lambda^2)} (t
x^*_{2,\uparrow} - \lambda y^*_{2,\downarrow}), \:
x^*_{4',\uparrow} = \frac{\epsilon_2^2 - t^2_h}{t_h(t^2+\lambda^2)} (t
x^*_{3',\uparrow} - \lambda y^*_{3',\downarrow}),
\nonumber\\
&&x^*_{7',\uparrow}=0, \: y^*_{1,\downarrow}=0, \: y^*_{7',\downarrow}=0, \:
y^*_{6,\downarrow} = \frac{t_f}{\epsilon_4} y^*_{2,\downarrow}, \:
y^*_{6',\downarrow} = \frac{t_f}{\epsilon_4} y^*_{3',\downarrow}, \: 
y^*_{5,\downarrow} = -y^*_{2,\downarrow}, \: y^*_{5',\downarrow} = - y^*_{3',\downarrow},
\nonumber\\
&&y^*_{3,\downarrow} = -\frac{\epsilon_2}{t_h} y^*_{2,\downarrow}, \:
y^*_{2',\downarrow} = -\frac{\epsilon_2}{t_h} y^*_{3',\downarrow}, \:
y^*_{4,\downarrow} = \frac{\epsilon_2^2 - t^2_h}{t_h(t^2+\lambda^2)} (\lambda
x^*_{2,\uparrow} + t  y^*_{2,\downarrow}),
\nonumber\\
&&y^*_{4',\downarrow} = \frac{\epsilon_2^2 - t^2_h}{t_h(t^2+\lambda^2)} (\lambda
x^*_{3',\uparrow} + t  y^*_{3',\downarrow}),
\label{E29}
\end{eqnarray}
where the $t^{\uparrow,\downarrow}=\lambda$ notation has been introduced, where
$\lambda$ represents the SOI coupling constant. From (\ref{E29}) is seen that
all coefficients of $\hat X^{\dagger}_{i}$ have been expressed in function of
$x^*_{2,\uparrow}, x^*_{3',\uparrow}, y^*_{2,\downarrow},y^*_{3',\downarrow}$. These can be
deduced based on the following relations
\begin{eqnarray}
x^*_{2,\uparrow}= \frac{-(\alpha \beta + \gamma \delta) x^*_{3',\uparrow} +
(\gamma \beta - \alpha \delta) y^*_{3',\downarrow}}{\alpha^2 + \gamma^2}, \:
y^*_{2,\downarrow}= \frac{-(\gamma \beta -\alpha \delta) x^*_{3',\uparrow} -
(\alpha \beta + \gamma \delta) y^*_{3',\downarrow}}{\alpha^2 + \gamma^2},         
\label{E30}
\end{eqnarray}  
where $\alpha = t a_{5,4} + \lambda b_{5,4}, \:
\beta = t a_{5,7} + \lambda b_{5,7}, \:
\gamma = t b_{5,4} - \lambda a_{5,4}, \:
\delta = t b_{5,7} - \lambda a_{5,7}$, and $x^*_{3',\uparrow}=p$, $y^*_{3',\downarrow}
=q$ are arbitrary parameters.

The physical meaning of these two free $p,q$ parameters can be given as follows:
given two set values of these parameters: $p_1,q_1$ and $p_2,q_2$, one can
construct two linearly independent $\hat X^{\dagger}_{i,1}$ and
$\hat X^{\dagger}_{i,2}$ solutions, which provide two linearly independent Hilbert
space vectors $|\Psi^0_1\rangle = \hat X^{\dagger}_{i,1}|0\rangle$, and
$|\Psi^0_2\rangle = \hat X^{\dagger}_{i,2}|0\rangle$, where $|0\rangle$ is the
vacuum state with no fermions present. These can be orthonormalized
based on the Gram-Schmidt procedure
\begin{eqnarray}
&&|\Psi_1\rangle = \frac{|\Psi^0_1\rangle}{\langle \Psi^0_1|\Psi^0_1 \rangle},
\quad |v\rangle = \langle \Psi^0_2 |\Psi_1 \rangle |\Psi_1\rangle,
\nonumber\\
&&|\Psi_2\rangle = \frac{|\Psi^0_2\rangle - |v\rangle}{(\langle \Psi^0_2| -
\langle v| )(|\Psi^0_2\rangle - |v\rangle)},  
\label{E30a}  
\end{eqnarray}
obtaining $\langle \Psi_1 | \Psi_2 \rangle = 0$ together with
$\langle \Psi_1 | \Psi_1 \rangle = \langle \Psi_2 | \Psi_2 \rangle = 1$.
Please note that for this to be done, two free parameters are needed, since
only one parameter, mathematically fixes the norm.

\subsubsection{Solution for $\hat X^{\dagger}_i$  at $\hat B \ne 0$}

The here presented solutions deduced in Appendix E occur at
\begin{eqnarray}
a_{5,7,d} a_{5,7,u} \ne b_{5,7} b'_{5,7}, \quad a_{5,4,d} a_{5,4,u} \ne b_{5,4} b'_{5,4},
\label{E31}  
\end{eqnarray}
where from the last four equations of (\ref{E28}) one has
$x_{1,\uparrow} = x_{7',\uparrow} = y_{1,\downarrow} = y_{7',\uparrow} =0$. The deduced
coefficients of the operator $\hat X^{\dagger}_i$ are as follows: The first 16
coefficients are given as
\begin{eqnarray}
&&x^*_{6,\uparrow}= \frac{t_f}{\epsilon_6-h} e^{-i(\phi_2+\phi_3)} x^*_{2,\uparrow}, \: 
y^*_{6,\downarrow}= \frac{t_f}{\epsilon_6+h} e^{-i(\phi_2+\phi_3)} y^*_{2,\downarrow},
\: x^*_{6',\uparrow}= \frac{t_f}{\epsilon_6-h} e^{i(\phi_2+\phi_3)} x^*_{3',\uparrow},
\nonumber\\
&&y^*_{6',\downarrow}= \frac{t_f}{\epsilon_6+h} e^{i(\phi_2+\phi_3)}y^*_{3',\downarrow},
\: x^*_{5,\uparrow}=- e^{-i(\phi_2+\phi_3)} x^*_{2,\uparrow}, \:  
y^*_{5,\downarrow}=- e^{-i(\phi_2+\phi_3)} y^*_{2,\uparrow}, \: 
x^*_{5',\uparrow}=- e^{i(\phi_2+\phi_3)} x^*_{3',\uparrow},
\nonumber\\
&&y^*_{5',\downarrow}=- e^{i(\phi_2+\phi_3)} y^*_{3',\downarrow}, \:
x^*_{3,\uparrow}=-e^{-i(2\phi_2 + 2\phi_3)}\frac{\epsilon_2 -h}{T_h} x^*_{2,\uparrow}, \:
y^*_{3,\downarrow}=-e^{-i(2\phi_2 + 2\phi_3)}\frac{\epsilon_2 +h}{T_h} y^*_{2,\downarrow},
\nonumber\\
&&x^*_{2',\uparrow}=-e^{i(2\phi_2 + 2\phi_3)}\frac{(\epsilon_2-h)(\epsilon_3-h)-t^2_h +
T_h(T_h +\epsilon_3 -h)}{T_h(\epsilon_3-h) +t^2_h} x^*_{3',\uparrow},
\nonumber\\
&&y^*_{2',\downarrow}=-e^{i(2\phi_2 + 2\phi_3)}\frac{(\epsilon_2+h)(\epsilon_3+h)-t^2_h +
  T_h(T_h +\epsilon_3 +h)}{T_h(\epsilon_3+h) +t^2_h} y^*_{3',\downarrow},
\nonumber\\
&&y^*_{4,\downarrow}= a x^*_{2,\uparrow} + b y^*_{2,\downarrow}, \:
x^*_{4,\uparrow}= c x^*_{2,\uparrow} + d y^*_{2,\downarrow}, \:
y^*_{4',\downarrow}= a' x^*_{3',\uparrow} + b' y^*_{3',\downarrow}, \:
x^*_{4',\uparrow}= c' x^*_{3',\uparrow} + d' y^*_{3',\downarrow},
\label{E32}
\end{eqnarray}
where one has for the presented prefactors
\begin{eqnarray}
&&a=-\lambda e^{-i(\phi_2+2\phi_3)} J_{-}, \: 
b=t e^{-i(\phi_2+2\phi_3)} J_{+}, \: 
c=t e^{-i(\phi_2+2\phi_3)} J_{-}, \: 
d=\lambda e^{-i(\phi_2+2\phi_3)} J_{+},
\nonumber\\
&&a'=\lambda e^{i(\phi_2+2\phi_3)} J_{-}, \:
b'=t e^{i(\phi_2+2\phi_3)} J_{+}, \: 
c'=t e^{i(\phi_2+2\phi_3)} J_{-}, \:
d'=-\lambda e^{i(\phi_2+2\phi_3)} J_{+},  
\nonumber\\
&&J_{+} = \frac{(T_h + \epsilon_2+h)[(\epsilon_2 +h)^2-t^2_h]}{(\lambda^2 +t^2)
[T_h(\epsilon_2 +h) +t^2_h)]}, \:
J_{-} = \frac{(T_h + \epsilon_2-h)[(\epsilon_2 -h)^2-t^2_h]}{(\lambda^2 +t^2)
[T_h(\epsilon_2 -h) +t^2_h)]}, 
\label{E33}
\end{eqnarray}
One notes that in (\ref{E33}) the $\epsilon_2=\epsilon_3$ equality has been used
satisfying the symmetry properties of the system. All presented coefficients
are expressed in function of the parameters $x^*_{2,\uparrow}, x^*_{3',\uparrow},
y^*_{2,\downarrow},y^*_{3',\downarrow}$. These are determined via
\begin{eqnarray}
&&x^*_{2,\uparrow} = \frac{Y'\beta' -X'\delta'}{\gamma' \beta' - \alpha'
\delta'}, \:
y^*_{2,\downarrow} = \frac{X'\gamma' -Y'\alpha'}{\gamma' \beta' - \alpha'
\delta'}, \: x^*_{3',\uparrow}=p', \: y^*_{3',\downarrow} = q',
\nonumber\\  
&&\alpha'= c a_{5,4,u}+ a b_{5,4}, \: \beta'= d a_{5,4,u} + b b_{5,4}, \:
\gamma' = c b'_{5,4} + a a_{5,4,d}, \: \delta' = d b'_{5,4} + b a_{5,4,d}, 
\nonumber\\
&&X'= - p'(a_{5,7,u} c' + b_{5,7} a') -q'( a_{5,7,u} d' + b_{5,7} b'),
\nonumber\\
&&Y'= - p'(a_{5,7,d} a' + b'_{5,7} c') -q'( a_{5,7,d} b' + b'_{5,7} d'), 
\label{E34}  
\end{eqnarray}
where $p',q'$ are arbitrary parameters. I note that the denominator present on
the first line of (\ref{E34}) is nonzero, since $\gamma' \beta' - \alpha'
\delta'= e^{-2i(\phi_2+2\phi_3)}J_{+}J_{-}(\lambda^2+t^2) (b_{5,4}b'_{5,4}-a_{5,4,d}
a_{5,4,u}) \ne 0$, see Eq.(\ref{E31}). Concerning $p',q'$ see the observation
below (\ref{E30}) relating the $p,q$ parameters.

\subsubsection{Other specific solutions for $\hat X^{\dagger}_i$}

In the Appendix F we present two more specific solutions for $\hat X^{\dagger}_i$,
one appearing i) in conditions when the requirement (\ref{E31}) is not
satisfied, and ii) the domain on which $\hat X^{\dagger}_i$ is defined, differs
from the domain presented in Fig.2. These solutions need supplementary
interdependences in between the parameters of the Hamiltonian (e.g. (\ref{E31})
taken as equality). These new interdependences not diminish substantially the
emergence possibilities of the phases described by the $\hat X^{\dagger}_i$
operators because e.g. the SOI coupling constants $\lambda,\lambda_c$ can be
continuously tuned by external electric fields \cite{Intr3}.

\subsubsection{The deduced exact ground states}

Once the $\hat X^{\dagger}_i$ operators have been obtained, we
deduce now the exact ground states for the Hamiltonian presented in
(\ref{Ez3}).

For this reason let us consider the wave vector
\begin{eqnarray}
|\Phi_1\rangle = \prod'_{i} \hat X^{\dagger}_i |0\rangle =
....\hat X^{\dagger}_i \hat X^{\dagger}_{i+2} \hat X^{\dagger}_{i+4}... |0\rangle   
\label{E35}
\end{eqnarray}
where $\prod'_i$ represents a product on each second lattice site, and
$\hat X^{\dagger}_i$ is taken from (\ref{E29},\ref{E30}) deduced at ${\bf B}=0$.
Note that $\hat X^{\dagger}_i$ places 1 electron on two nearest neighbor
pentagons excepting their left and right end sites (sites 1 and 7' in Fig.2).
Inside the block covered by $\hat X^{\dagger}_i$ (sites 2,3,4,5,6,4',2',3',5',6'
in Fig.2), there are not present two electrons (i.e. double occupancy is not
present), hence $\hat H_U |\Phi_1\rangle =0$ holds. Furthermore, inside this
block there are not present two nearest neighboring sites occupied both at
least once, so $\hat H_V |\Psi_1\rangle =0$ also holds. In between two nearest
neighbor $\hat X^{\dagger}_i$ and $\hat X^{\dagger}_{i+2}$ operators two empty sites
are present (the horizontal bonds starting to the right from the site 7', and
starting to the left from the site 1 in Fig.2), so on these sites
$\hat H_U |\Phi_1\rangle =0$, and $\hat H_V |\Phi_1\rangle =0$ are also
satisfied, consequently
\begin{eqnarray}
(\hat H_U + \hat H_V)|\Phi_1\rangle =0
\label{E36}
\end{eqnarray}
is satisfied. But $\hat X^{\dagger}_i$  was deduced from the
condition (\ref{E26}),
so (\ref{E36}) type of relation is satisfied also for the first term of the
Hamiltonian (\ref{Ez3}), as a consequence $\hat H |\Phi_1\rangle =0$ holds. But
since $\hat H$ is build up only from positive semidefinite terms, zero is the
smallest possible eigenvalue of $\hat H$. As a consequence $|\Phi_1\rangle =
|\Phi_{g,1}\rangle$ is the ground state of $\hat H$ at a number of $N=N_c/2$
particles in the system. The system has totally $6N_c$ sites, so the maximum
number of electrons in the system is $N_{max}=12 N_c$, consequently $N_c/2$
carriers represent $N/N_{max}= 1/24$ particle concentration (quarter filled
lowest band). This concentration value is placed below system half filling
($6N_c$ electrons), that is why is called ``ground state in the low density''
case. Physically this state represents a charge density wave, since on all
second horizontal bonds (e.g. bond 7' to ${\bf i}+2{\bf a}$ in Fig.2) carriers
are missing. This state is insulating. For the solution to occur, from the
point of view of the two-particle interaction terms, $U_n > 0$, $V > 0$, but
otherwise arbitrary are needed.

One more problem must be treated, namely the interplay between $|\Phi_1\rangle$
mentioned above, and $|\Psi_{\alpha}\rangle$, $\alpha=1,2$ presented in
(\ref{E30a}). For this to be understood, we must turn back to (\ref{E9}), which
provide the solution of the matching equations used here, together with the
conditions (\ref{E10}), these last providing the parameter space region where
the solution is valid. Checking the conditions (\ref{E10}) for their meaning in
the band structure [see Ref. \cite{Intr3}], one sees, that it provides a lowest
flat band, which is double degenerate. Consequently, in $|\Phi_{g,1}\rangle$ one
can use at the sites i or $\hat X^{\dagger}_{i,1}$ or $\hat X^{\dagger}_{i,2}$,
but not both, maintaining the quarter filling of the lowest band.

I must also underline that a ground state solution of the type (\ref{E35})
holds as well bellow $c_0 = N/N_{max}= 1/24$ system filling. In this case
\begin{eqnarray}
|\Phi_{g,1}(c < c_0)\rangle = \prod_{i \in D} \hat X^{\dagger}_i |0\rangle, 
\label{E37}  
\end{eqnarray}
where $D$ is a lattice sites manifold containing $N < c_0 N_{max}$ lattice sites
$D={i_1,i_2,...i_N}$, where the sites contained in the $D$ domain must be at
least at a distance $2{\bf a}$ from each other. In this case the charge density
wave nature of the ground state is lost, and the ground state is built up from
isolated clusters. The insulating nature of the system is maintained. 

Now let us consider the ${\bf B} \ne 0$ case. In this situation one works
with $\hat X^{\dagger}_i$ deduced in (\ref{E32}-\ref{E34}), and we consider here
the $\hat H_Z=0$ case only, so only the effect of the external magnetic field
on the orbital motion of electrons is considered described by the Peierls phase
factors. In this case, at $c_0=1/24$ filling the ground state expression remains
that given in (\ref{E35}), and at $c < c_0$ filling, that given in (\ref{E37}).
But the situation is that the solution of the matching equations is satisfied
only for fixed $\phi$ values (see Ref.\cite{Intr3}). Hence when the magnetic
field is switched on, the charge density wave phase disappears, and reappears
at a fixed external magnetic field value. This is an interesting example which
shows how a static magnetic fields can influence a static electric charge
distribution. Furthermore, if (\ref{E35}) and (\ref{E37}) are not allowed
solutions at $\phi \ne 0$, the lowest band will be dispersive, and the system
becomes conducting. This provides a nice example demonstrating that a static
magnetic field is able to turn an insulating phase to a conducting phase. The
presented examples may have broad application possibilities in leading
technologies.

When the conditions presented in the Appendix F.1 are satisfied, the expression
of the ground state wave vectors from (\ref{E35}) and (\ref{E37}) remain valid
at $V=0$ only, because in this case, along the horizontal bonds connecting the
unit cells, two single occupied nearest neighbor sites can be present in the
system. Because in this case also the sites 1 and 7' in Fig.2 could be occupied,
also the charge density wave nature of the ground state is lost.

When the conditions presented in the Appendix F.2 are present, the
$\hat X^{\dagger}_i$ operator acts only on the four middle sites in each
unit cell (2,3,5,6 in Fig.2). Now the (\ref{E35}), and (\ref{E37}) ground state
expressions remain valid also at $V \ne 0$, and could contain even $N_c$
operators, so in this case the ground state can be written up to a carrier
concentration $N/N_{max}=1/12$. One unit cell will be ferromagnetic, but the
cells will be uncorrelated, so the ground state will be paramagnetic. A
possible exception from this rule is provided by the situation when the lowest
two bands touch each other: this situation will provide a ferromagnetic state
based on a similar mechanism as described in Ref. \cite{Intr16}.

\subsection{Ground states in the high density case.}

The Hamiltonian one analyses now is presented in (\ref{Ez8}) used with
periodic boundary conditions. The ground state construction now begins with
\begin{eqnarray}
|\Phi'_1\rangle = \prod^{N_c}_{i=1} \prod^5_{z=1} \prod_{\sigma=\uparrow,\downarrow}
\hat B^{\dagger}_{i,z,\sigma} |0\rangle.
\label{EEE39}
\end{eqnarray}
This strategy is based on the fact that
$\hat B^{\dagger}_{i,z,\sigma}\hat B^{\dagger}_{i,z,\sigma}=0$, hence the first term
from (\ref{Ez8}) satisfies
\begin{eqnarray}
(\sum_{i,z,\sigma} \hat B_{i,z,\sigma} \hat B^{\dagger}_{i,z,\sigma})|\Phi'_1\rangle = 0.
\label{EEE40}
\end{eqnarray}  
The problem to be further solved now, is to introduce $|\Phi'_1\rangle$ in the
kernel of the remaining positive semidefinite operators from the Hamiltonian,
i.e. $\sum_i \sum_{n=1,2,..6} U_n\hat P_{i,n}$ and $\sum_{<l,k>} V \hat R_{<l,k>}$.
The first term from these two requires at least one
electron on each site. The study of (\ref{EEE39}) shows that $|\Phi_1\rangle$
already introduces in the system $10N_c$ electrons, and with periodic boundary
conditions this can leave one empty site per unit cell. So in order to do not
have empty sites in the system, we must multiply the operators present in
$|\Phi_1\rangle$ by a term
\begin{eqnarray}
\hat B^{\dagger} = \prod_i c^{\dagger}_{i,n(i),\sigma(i)}(1-\hat n_{i,n(i),-\sigma(i)}),
\label{EEE41}
\end{eqnarray}  
which introduces in each unit cell one electron with an arbitrary spin
$\sigma(i)$ on an arbitrary site $n(i)$, which is empty. The $\hat B^{\dagger}$
operator
from (\ref{EEE41}) presents another novelty in comparison to its previous
applications (see e.g. Ref. \cite{Intr8,Intr9,Intr10}). Namely, together with
the operators present in (\ref{EEE39}), preserves in the presented system only
sites containing at least one electron, but supplementary such that two
single occupied sites cannot be in nearest neighbor positions. 
Hence, the wave vector
\begin{eqnarray}
|\Phi_1\rangle = ( \prod^{N_c}_{i=1} \prod^5_{z=1} \prod_{\sigma=\uparrow,\downarrow}
\hat B^{\dagger}_{i,z,\sigma}) \hat B^{\dagger} |0\rangle,
\label{EEE42}
\end{eqnarray}
has at least one electron on each site, consequently
\begin{eqnarray}
(\sum_i \sum_{n=1,2,..6} U_n\hat P_{i,n})|\Phi_1\rangle =0,
\label{EEE43}
\end{eqnarray}
holds. But besides this, one also has
\begin{eqnarray}
(\sum_{<l,k>} V \hat R_{<l,k>} )|\Phi_1\rangle =0,
\label{EEE43a1}
\end{eqnarray}
because $|\Phi_1\rangle$ not only contains sites at least once occupied, but
supplementary not contains single occupied states in nearest neighbor positions.
Since $|\Phi_1\rangle$ also satisfies (\ref{EEE40}), so $(\hat H -C)|\Phi_1
\rangle =0$, hence $|\Phi_1\rangle$ is the ground state of the system, i.e.
$|\Phi_{g,1}\rangle =|\Phi_1\rangle$, and the ground state energy is $E_g=C$.  
Since now $N/N_{max} > 5/6$ is above system half filling ($N/N_{max}=1/2$),
the here presented ground state is called high density ground state.

The $\hat B_{i,z,\sigma}$ operators present in (\ref{EEE42}) are those defined in
(\ref{Ez2}), the matching equations remain at ${\bf B}=0$ the relations
(\ref{Ez4}-\ref{Ez7}) and at ${\bf B} \ne 0$ the equations
(\ref{E13}-\ref{E16}), but with the substitutions presented in (\ref{A8}).
Consequently, now the matching equations contain all coupling constants,
including those present in SOI ($\lambda,\lambda_c$), in $\hat H_U$, and
$\hat H_V$.

At ${\bf B}=0$ using for the block operator coefficients of $\hat B_{i,z,\sigma}$
deduced from (\ref{Ez4}-\ref{Ez7}), given by the conditions from (\ref{E10}),
one finds again an effective flat band which is created not by the bare band
structure, but by the $U,V$ interaction terms. The ground state
$|\Phi_{g,1}\rangle$ presented in (\ref{EEE42}) is nonmagnetic, and corresponds
to $c_1=N/N_{max}=11/12$ filling (upper band half filled). Turning on the
external magnetic field, the mathematical expression of the deduced ground state
remains the expression (\ref{EEE42}), but now the block operator parameters must
be taken from the matching equations solution presented in (\ref{E17}-
\ref{E24}).

The ground state can be obtained also in the region $c_1 < c \leq 1$. For this,
in the case of the carrier concentration $c=(N_1 + N')/N_{max}$, where
$N_1= c_1 N_{max}$, and $ 1 \leq N' \leq N_c)$, a supplementary operator
$\hat B^{\dagger}_c = \prod_{j=1}^{N'} \hat c^{\dagger}_{k_j,\sigma_j}$ must be added to
the operators present in (\ref{EEE42}), where $k_j,\sigma_j$ are arbitrary
momenta and spin projections. The ground state becomes in this case of the form
\begin{eqnarray}
|\Phi_{g,1} (c > c_1) \rangle = ( \prod^{N_c}_{i=1} \prod^5_{z=1} \prod_{
\sigma=\uparrow,\downarrow} \hat B^{\dagger}_{i,z,\sigma}) \hat B^{\dagger}
\hat B^{\dagger}_c |0\rangle,
\label{EEE45}
\end{eqnarray}  
which is a nonmagnetic and itinerant ground state.

\section{Summary and conclusions}

The paper analyses a pentagon chain type of conducting polymer in the presence
of spin-orbit interactions SOI ($\lambda,\lambda_c$), on-site (U) and nearest
neighbor (V) Coulomb repulsions, in the presence of external fields
${\bf E},{\bf B}$. Both fields are applied perpendicular to the plane containing
the chain. ${\bf E}$ continuously tunes the SOI interaction strength, while
the action of ${\bf B}$ is taken into account both at the level of the orbital
motion of electrons (Peierls phase factors), and its action on the spin degrees
of freedom (Zeeman term). Usually $\lambda << U$ holds. Besides, if
the SOI interaction strength is even small, it has essential effects on the
physical properties of the system, since breaks the spin projection double
degeneracy of each band. In these conditions, the perturbative description
of the system in both small and high coupling constant limits is questionable,
hence exact methods are used in the study. On the other side this job has its
difficulties because the system is nonintegrable, so special technique must be
applied in order to follow this route. On this line one uses the method based
on positive semidefinite operator properties, which provides the first exact
ground states for conducting polymers in the presence of SOI. Furthermore, for
the first time this exact procedure is used in the case of conducting polymers
in the presence of both U, and  V Coulomb interactions. Besides, in this
technique also for the first time the common action of ${\bf B}$ on spin and
orbital degrees of freedom is considered. In these conditions the paper
concentrates on the development of the technique used in the presented
conditions: the transformation of the Hamiltonian in positive
semidefinite form, the solution technique of the matching equations, the
construction strategy of the exact ground states. The study of the physical
properties of the deduced ground states is only sketched, since this job
requires considerable future work, hence the detailed presentation of the
physical properties of all deduced phases is left for a future publication.
Even in these conditions, the presented physical properties show a broad
spectrum of interesting physical characteristics as e.g. emerging charge
density wave phases, modification possibility of a static charge distribution
by an external static magnetic field, transition possibilities between
insulating and conducting phases provided by external magnetic fields,
emergence of effective flat bands provided by the two-particle local or
non-local Coulomb interaction terms, peculiar ferromagnetic states, etc.

\appendix

\section{Another transformation in positive semidefinite form}

We wrote the system Hamiltonian from Eq.(\ref{Ez1}) in the form
$\hat H = \hat H_K + \hat H_U + \hat H_V$. First we concentrate on the
transformation of the $\hat H_U$ Hubbard term. It can be observed that
\begin{eqnarray}
\hat H_U &=& \sum_i \sum_{n=1,2,..6} U_n \hat n_{i,n}^{\uparrow} 
\hat n_{i,n}^{\downarrow} = \sum_i \sum_{n=1,2,..6} U_n [\hat n_{i,n}^{\uparrow}
\hat n_{i,n}^{\downarrow}-(\hat n_{i,n}^{\uparrow}+\hat n_{i,n}^{\downarrow})+1]
\nonumber\\
&+&\sum_i \sum_{n=1,2,..6} U_n (\hat n_{i,n}^{\uparrow}+\hat n_{i,n}^{\uparrow}) -
\sum_i \sum_{n=1,2,..6} U_n = \sum_i \sum_{n=1,2,..6} U_n\hat P_{i,n}
\nonumber\\
&+&\sum_i \sum_{n=1,2,..6} U_n (\hat n_{i,n}^{\uparrow}+\hat n_{i,n}^{\downarrow}) -
N_c  \sum_{n=1,2,..6} U_n,
\label{A1}
\end{eqnarray}
where $N_c$ represents the number of lattice sites (cells), while
$\hat P_{i,n} =[\hat n_{i,n}^{\uparrow}
\hat n_{i,n}^{\downarrow}-(\hat n_{i,n}^{\uparrow}+\hat n_{i,n}^{\uparrow})+1]$
is a positive semidefinite operator which provides its minimum possible
eigenvalue zero, when one has at least one electron on the $(i,n)$ site.

After this step one transforms the $\hat H_V$ term. For this, we introduce the
site notation $(l,k)$ which cover all sites, not only the lattice sites. In
these conditions, denoting the nearest neighbor $l,k$ sites (which are taken
into consideration only once) by $<l,k>$, one has
\begin{eqnarray}
\hat H_V &=& V \sum_{<k,l>} \hat n_l \hat n_k = V \sum_{<l,k>}[\hat n_l \hat n_k -
2(\hat n_l + \hat n_k) + 4] + 2 V \sum_{<l,k>}(\hat n_l + \hat n_k)
\nonumber\\
&-&4\sum_{<l,k>} V = \sum_{<l,k>} V \hat R_{<l,k>} + \sum_i \sum_{n=1,2,..6}
2V r_n \hat n_{i,n} - 28N_c V
\label{A1a1}  
\end{eqnarray}
where we have used the fact that 7 bonds are present in an unit cell, so
$\sum_{<l,k>} = 7 \sum_i$ where i runs over the lattice sites, furthermore
$\sum_{<l,k>}(\hat n_l + \hat n_k) = \sum_i \sum_{n=1,2,..6} r_n \hat n_{i,n}$,
where the coefficients $r_1=3,r_2=2,r_3=2,r_4=3,r_5=3,r_6=1$ show how many time
a given site in an unit cell appears in the $\sum_{<l,k>}$ sum. Here the
$\hat R_{<l,k>}= [\hat n_l \hat n_k - 2(\hat n_l + \hat n_k) + 4]$ is a positive
semidefinite operator which provides its minimum possible eigenvalue zero, when
the $l,k$ nearest neighbor sites are at least once occupied, but such that
at least one of them is doubly occupied.

Now we introduce in the Hamiltonian $\hat H$ from (\ref{Ez1}) the $\hat H_U$
term from (\ref{A1}) and the $\hat H_V$ term from (\ref{A1a1}) obtaining
\begin{eqnarray}
\hat H &=& \hat H_K + (\sum_i \sum_{n=1,2,..6} U_n \hat P_{i,n} +
\sum_i \sum_{n=1,2,..6} U_n (\hat n_{i,n}^{\uparrow}+\hat n_{i,n}^{\downarrow}) -
N_c  \sum_{n=1,2,..6} U_n)
\nonumber\\
&+& (\sum_{<l,k>} V \hat R_{<l,k>} + \sum_i \sum_{n=1,2,..6}
2V r_n \hat n_{i,n} - 28N_c V)  
\label{A2}
\end{eqnarray}
This Hamiltonian from Eq.(\ref{A2}) will be written in the positive
semidefinite form
\begin{eqnarray}
\hat H =\sum_{i,z,\sigma} \hat B'_{i,z,\sigma} \hat B'^{\dagger}_{i,z,\sigma} +
\sum_i \sum_{n=1,2,..6} U_n \hat P_{i,n} + \sum_{<l,k>} V \hat R_{<l,k>} + C  
\label{A3}
\end{eqnarray}
where $C$ is a constant, and $\hat B'_{i,z,\sigma}$ has exactly the form
presented in Eq.(\ref{Ez2}) with new coefficients. Denoting the anticommutators
$\{\hat B'_{i,z,\sigma},\hat B'^{\dagger}_{i,z,\sigma}\} = y_{z,\sigma}$ (note that
$y_{z,\sigma}$ is the same in each cell placed at the lattice site i),
Eq.(\ref{A3}) becomes of the form
\begin{eqnarray}
\hat H =- \sum_{i,z,\sigma} \hat B'^{\dagger}_{i,z,\sigma} \hat B'_{i,z,\sigma} + N_c
\sum_{z,\sigma}y_{z,\sigma} + \sum_i \sum_{n=1,2,..6} U \hat P_{i,n} +
\sum_{<l,k>} V \hat R_{<l,k>} + C  
\label{A4}
\end{eqnarray}
Now since the equations (\ref{A2}) and (\ref{A4}) contain the same Hamiltonian
in the left side, equating the right sides on finds
\begin{eqnarray}
&&\hat H_K + \sum_{i,\sigma} (U_n+2Vr_n) \hat n_{i,n}^{\sigma} - q_U \sum_{i,\sigma}
\hat n_{i,n}^{\sigma}
+ q_U N -N_c  \sum_{n=1,2,..6} U_n - 28N_c V =
\nonumber\\
&&- \sum_{i,z,\sigma} \hat B'^{\dagger}_{i,z,\sigma}
\hat B'_{i,z,\sigma} + N_c \sum_{z,\sigma}y_{z,\sigma} + C  
\label{A5}
\end{eqnarray}
where $N$ represents the number of electrons, and $q_U$ is a constant (note
that we added to the left side $-q_U N + q_U N =0$). 
Now we deduce the block operators $\hat B'_{i,z,\sigma}$ based on the
transformation
\begin{eqnarray}
\sum_{i,z,\sigma} \hat B'^{\dagger}_{i,z,\sigma}\hat B'_{i,z,\sigma} =
-[\hat H_K + \sum_{i,n,\sigma} (U_n + 2Vr_n) \hat n_{i,n}^{\sigma} - q_U \sum_{i,\sigma}
\hat n_{i,n}^{\sigma}]
\label{A6}
\end{eqnarray}
which taken into account in Eq.(\ref{A5}) provides for the constant $C$ the
value
\begin{eqnarray}
C=q_U N - N_c \sum_{z,\sigma}y_{z,\sigma} - N_c \sum_{n=1,2,..6} U_n -28 V N_c
\label{A7}
\end{eqnarray}
In conclusions, based on (\ref{A3}) the starting Hamiltonian from (\ref{Ez1})
can be transformed in the positive semidefinite form (\ref{A3}), where the
constant $C$ is given in (\ref{A7}), and the block operator coefficient must
be deduced based on (\ref{A6}). 

But one realizes that the block operators $\hat B'_{i,z,\sigma}$ are exactly the
block operators $\hat B_{i,z,\sigma}$ from (\ref{Ez2}) whose matching equations
are presented in Eqs.(\ref{Ez4}-\ref{Ez7}) with the condition that at the level
of the Hamiltonian parameters the following changes are made
\begin{eqnarray}
t^{\sigma,\sigma}_{i,j} \to -t^{\sigma,\sigma}_{i,j}, \quad 
t^{\sigma,\sigma'}_{i,j} \to -t^{\sigma,\sigma'}_{i,j}, \quad
\epsilon^{\sigma,\sigma}_i \to q_U -( \epsilon^{\sigma,\sigma}_i +U_i +2Vr_i), \quad
\epsilon^{\sigma,\sigma'}  \to - \epsilon^{\sigma,\sigma'}
\label{A8}
\end{eqnarray}
where $\sigma' \ne \sigma$ holds. Hence with the changes from Eq.(\ref{A8})
the Hamiltonian from (\ref{Ez1}) becomes of the form
\begin{eqnarray}
\hat H =\sum_{i,z,\sigma} \hat B_{i,z,\sigma} \hat B^{\dagger}_{i,z,\sigma} +
\sum_i \sum_{n=1,2,..6} U_n \hat P_{i,n} + \sum_{<l,k>} V \hat R_{<l,k>} + C,  
\label{A9}
\end{eqnarray}
where the constant $C$ is given in (\ref{A7}), and the matching equations for
the $\hat B_{i,z,\sigma}$ block operators defined in (\ref{Ez2}) remain does
presented in Eqs.(\ref{Ez4}-\ref{Ez7}) at ${\bf B}=0$ and Eqs.(\ref{E13}-
\ref{E16}) at ${\bf B} \ne 0$ conditioned by the interchanges
described in (\ref{A8}). 

\section{Solution of the matching equations at $B = 0$}

\subsubsection{The first ten equations.}

One notes that during this solution one uses $t=t^*$, and concentrates on
Eqs.(\ref{Ez4}). Since a solution of the matching equations is presented here
for the first time in the paper, more details are mentioned during the
presentation.

From the rows 3-8 of (\ref{Ez4}), six $a_{i,j}$ coefficients can be expressed in
function of other $a_{i',j'}$ coefficients, reducing in this manner the number
of unknown variables by six:
\begin{eqnarray}
&&a_{1,5}=\frac{t}{a^*_{1,1}}, \: a^*_{1,2}=\frac{t}{a_{1,1}}, \:
a_{3,3}=\frac{t}{a^*_{3,4}}, \: a^*_{3,5}=\frac{t}{a_{3,4}}, \: 
a^*_{2,3}=\frac{t_h}{a_{2,2}}, \: a_{4,5}=\frac{t_f}{a^*_{4,6}},
\nonumber\\
&&a_{1,2}=a_{1,5}, \quad a_{3,3}=a_{3,5}.  
\label{Ez9}
\end{eqnarray}
Using these relations, the remaining four equalities of (\ref{Ez4})
provide  
\begin{eqnarray}
t_h > 0.
\label{Ez10}
\end{eqnarray}
Indeed
\begin{eqnarray}
&&a_{2,5}=-\frac{t^2}{a^*_{2,2}|a_{1,1}|^2},
\nonumber\\
&&t_h=\frac{|a_{2,2}|^2|a_{1,1}|^2}{|a_{3,4}|^2}, \: => \:
|a_{3,4}|^2=\frac{|a_{2,2}|^2|a_{1,1}|^2}{t_h},
\nonumber\\
&&t_c = a^*_{5,7}a_{5,4}+ b^*_{5,7}b_{5,4},
\nonumber\\
&&b^*_{5,7}b_{5,4}={b'}^*_{5,7}b'_{5,4},
\label{Ez11}
\end{eqnarray}
The here obtained first two relations are such obtained, that one keeps the
first equality of (\ref{Ez4}), and one divides the first two equalities of
(\ref{Ez4}).

\subsubsection{The second ten equations.}

In what will follows on concentrates to Eqs.(\ref{Ez5}).
The procedure used here is as follows:
From consecutive two equations of (\ref{Ez5}) we form pairs. From the
first component of the pair we express the $b$ coefficient. After this step
by dividing the two components of the pair, one connects $b'$ to $b$. In this
manner, the following ten equations are obtained:
\begin{eqnarray}
&&{b'}^*_{1,1}= \frac{t^{\uparrow,\downarrow}_{1,5}}{a_{1,5}},
\nonumber\\
&&{b'}^*_{1,1}=\alpha b^*_{1,1}
\nonumber\\
&&b_{1,1}=\frac{t^{\uparrow,\downarrow}_{2,1}}{a^*_{1,2}},
\nonumber\\
&&b_{1,1}=\alpha b'_{1,1}
\nonumber\\
&&b_{3,4}=\frac{t^{\uparrow,\downarrow}_{5,4}}{a^*_{3,5}},
\nonumber\\
&&b_{3,4}=\alpha b'_{3,4}
\nonumber\\
&&{b'}^*_{3,4}= \frac{t^{\uparrow,\downarrow}_{4,3}}{a_{3,3}},
\nonumber\\
&&{b'}^*_{3,4}=\alpha b^*_{3,4}
\nonumber\\
&&t^{\uparrow,\downarrow}_c={b'}^*_{5,7}a_{5,4}+a^*_{5,7}b_{5,4},
\nonumber\\
&&\alpha_c= \frac{{b'}^*_{5,7}a_{5,4}+a^*_{5,7}b_{5,4}}{{b}^*_{5,7}a_{5,4}+a^*_{5,7}
{b'}_{5,4}},
\label{Ez12}
\end{eqnarray} 
Now we divide the second and fourth line, and similarly the sixth and eight
line respectively, obtaining
\begin{eqnarray}
|b'_{1,1}|^2 = |b_{1,1}|^2, \quad |b'_{3,4}|^2 = |b_{3,4}|^2, \: => \: |\alpha|^2=1
\label{Ez13}
\end{eqnarray}
Introducing now the phase factors $\phi_{\alpha}, \phi_1, \phi_2$, based on
(\ref{Ez13}), the following relations are obtained:
\begin{eqnarray}
b'_{1,1}= e^{i\phi_1} b_{1,1}, \quad b'_{3,4}= e^{i\phi_2} b_{3,4}, \quad
\alpha=e^{i\phi_{\alpha}},
\label{Ez14}
\end{eqnarray}
But, from the fourth and sixth line of (\ref{Ez12}) one has $1=e^{i\phi_{\alpha}}
e^{i\phi_1}$, and  $1=e^{i\phi_{\alpha}}e^{i\phi_2}$ respectively, so
\begin{eqnarray}
e^{i\phi_1}=e^{-i\phi_{\alpha}}, \quad e^{i\phi_2}=e^{-i\phi_{\alpha}},
\label{Ez15}
\end{eqnarray}
consequently, from (\ref{Ez14}) one obtains
\begin{eqnarray}
b'_{1,1}= e^{-i\phi_{\alpha}} b_{1,1}, \quad b'_{3,4}= e^{-i\phi_{\alpha}} b_{3,4}.
\label{Ez16}
\end{eqnarray}
I further note that from (\ref{Ez12}), we further remain with the following
relations
\begin{eqnarray}
&&b_{1,1}=\frac{t^{\uparrow,\downarrow}_{2,1}}{a^*_{1,2}}= 
\frac{{t^{\uparrow,\downarrow}_{1,5}}^*}{a^*_{1,5}}e^{i\phi_{\alpha}},
\nonumber\\
&&b_{3,4}=\frac{t^{\uparrow,\downarrow}_{5,4}}{a^*_{3,5}}= 
\frac{{t^{\uparrow,\downarrow}_{4,3}}^*}{a^*_{3,3}}e^{i\phi_{\alpha}},
\nonumber\\
&&t^{\uparrow,\downarrow}_c={b'}^*_{5,7}a_{5,4}+a^*_{5,7}b_{5,4},
\nonumber\\
&&\alpha_c= \frac{{b'}^*_{5,7}a_{5,4}+a^*_{5,7}b_{5,4}}{{b}^*_{5,7}a_{5,4}+a^*_{5,7}
{b'}_{5,4}}.
\label{Ez17}
\end{eqnarray}
Based on the second line of (\ref{Ez8}) we further have:
\begin{eqnarray}
t^{\uparrow,\downarrow}_{2,1}={t^{\uparrow,\downarrow}_{1,5}}^*e^{i\phi_{\alpha}}, \quad
t^{\uparrow,\downarrow}_{5,4}={t^{\uparrow,\downarrow}_{4,3}}^*e^{i\phi_{\alpha}}, 
\label{Ez18}
\end{eqnarray}
and what still remains to be analyzed from the second group of ten equations
can be summarized as follows
\begin{eqnarray}
&&b_{1,1}=\frac{t^{\uparrow,\downarrow}_{2,1}}{a^*_{1,2}},
\nonumber\\
&&b_{3,4}=\frac{t^{\uparrow,\downarrow}_{5,4}}{a^*_{3,5}},
\nonumber\\
&&t^{\uparrow,\downarrow}_c={b'}^*_{5,7}a_{5,4}+a^*_{5,7}b_{5,4},
\nonumber\\
&&\alpha_c= \frac{{b'}^*_{5,7}a_{5,4}+a^*_{5,7}b_{5,4}}{{b}^*_{5,7}a_{5,4}+a^*_{5,7}
{b'}_{5,4}},
\label{Ez19}
\end{eqnarray}

\subsubsection{The ten equations provided by the on-site one particle
  potentials.}

\subsubsection*{3.a) The relations relating the b coefficients.}

Taking into account that $\epsilon_1^{\uparrow,\uparrow}=
\epsilon_1^{\downarrow,\downarrow}$, and $\epsilon_4^{\uparrow,\uparrow}=
\epsilon_4^{\downarrow,\downarrow}$, based on the 5-6 and 7-8 equalities of
(\ref{Ez6}), one finds
\begin{eqnarray}
&&|b'_{1,1}|^2 +|b'_{5,7}|^2=|b_{1,1}|^2 +|b_{5,7}|^2,
\nonumber\\
&&|b'_{3,4}|^2 +|b'_{5,4}|^2=|b_{3,4}|^2 +|b_{5,4}|^2,
\label{Ez20}
\end{eqnarray}
so taking into account (\ref{Ez13}), we find
\begin{eqnarray}
|b'_{5,7}|^2=|b_{5,7}|^2, \quad |b'_{5,4}|^2=|b_{5,4}|^2.
\label{Ez21}
\end{eqnarray}
Now as in the case of (\ref{Ez14}), we introduce
\begin{eqnarray}
b'_{5,4}=e^{i\phi_3}b_{5,4}, \quad b'_{5,7}=e^{i\phi_4}b_{5,7}.
\label{Ez22}
\end{eqnarray}
Introducing these results in the last equality of (\ref{Ez11}), one finds
\begin{eqnarray}
b^*_{5,7}b_{5,4}=e^{-i\phi_4}b^*_{5,7}e^{i\phi_3}b_{5,4}, \: => \: e^{i\phi_3}=e^{i\phi_4}.
\label{Ez23}
\end{eqnarray}
Introducing these results in the last equality of (\ref{Ez19}), one obtains
\begin{eqnarray}
\alpha_c=\frac{e^{-i\phi_3}b^*_{5,7} a_{5,4}+ a^*_{5,7}b_{5,4}}{
b^*_{5,7} a_{5,4}+ a^*_{5,7}b_{5,4}e^{i\phi_3}}, \: => \:
\alpha_c=e^{-i\phi_3} \: => \: \alpha_c =e^{i\phi_{\alpha_c}}=e^{-i\phi_3}.
\label{Ez24}
\end{eqnarray}
After this step, using $\epsilon^{\sigma,-\sigma}_i=0, \: i=1,4$, the first two
equalities of (\ref{Ez7}) provide information relating the $b_{5,4}$ and
$b_{5,7}$ coefficients as follows
\begin{eqnarray}
&&(a^*_{5,7}b_{5,7}+b^*_{5,7}a_{5,7}e^{i\phi_{\alpha_c}})=-(a^*_{1,1}b_{1,1}+b^*_{1,1}a_{1,1}
e^{i\phi_{\alpha}}),
\nonumber\\
&&(a^*_{5,4}b_{5,4}+b^*_{5,4}a_{5,4}e^{i\phi_{\alpha_c}})=-(a^*_{3,4}b_{3,4}+b^*_{3,4}a_{3,4}
e^{i\phi_{\alpha}}).
\label{Ez25}
\end{eqnarray} 
The obtained equalities (\ref{Ez25}) have an interesting structure, E.g.
the first equality and its complex conjugate provide
\begin{eqnarray}
&&(a^*_{5,7}b_{5,7}+b^*_{5,7}a_{5,7}e^{i\phi_{\alpha_c}})=-(a^*_{1,1}b_{1,1}+b^*_{1,1}a_{1,1}
e^{i\phi_{\alpha}}),
\nonumber\\
&&(a_{5,7}b^*_{5,7}+b_{5,7}a^*_{5,7}e^{-i\phi_{\alpha_c}})=-(a_{1,1}b^*_{1,1}+b_{1,1}a^*_{1,1}
e^{-i\phi_{\alpha}}) \: => 
\nonumber\\
&&e^{-i\phi_{\alpha_c}}(a_{5,7}b^*_{5,7}e^{i\phi_{\alpha_c}}+b_{5,7}a^*_{5,7})=-
e^{-i\phi_{\alpha}}(a_{1,1}b^*_{1,1}e^{i\phi_{\alpha}}+b_{1,1}a^*_{1,1}).
\label{Ez26}
\end{eqnarray}
In this case based on the first row of (\ref{Ez26}), the third row will satisfy
\begin{eqnarray}
e^{-i\phi_{\alpha_c}}=e^{-i\phi_{\alpha}}, \: => \:
e^{i\phi_{\alpha_c}}=e^{i\phi_{\alpha}}.
\label{Ez27}
\end{eqnarray}
Similar result is obtained from the second equality of (\ref{Ez25}),
hence (\ref{Ez25}) can be written as
\begin{eqnarray}
&&(a^*_{5,7}b_{5,7}+b^*_{5,7}a_{5,7}e^{i\phi_{\alpha}})=-(a^*_{1,1}b_{1,1}+b^*_{1,1}a_{1,1}
e^{i\phi_{\alpha}}),
\nonumber\\
&&(a^*_{5,4}b_{5,4}+b^*_{5,4}a_{5,4}e^{i\phi_{\alpha}})=-(a^*_{3,4}b_{3,4}+b^*_{3,4}a_{3,4}
e^{i\phi_{\alpha}}).
\label{Ez28}
\end{eqnarray}

\subsubsection*{3.b) The relations relating the a coefficients}

Now one analyses the equalities 3-4 of (\ref{Ez6}) in which we use as well
(\ref{Ez9}) obtaining
\begin{eqnarray}
&&\epsilon_2 = \frac{t^2}{|a_{1,1}|^2} + |a_{2,2}|^2,
\nonumber\\
&&\epsilon_2 = \frac{t_h^2}{|a_{2,2}|^2} + \frac{t^2}{|a_{3,4}|^2} 
, \: => \:
\epsilon_2 = \frac{t_h^2}{|a_{2,2}|^2} + \frac{t^2t_h}{|a_{1,1}|^2 |a_{2,2}|^2} 
\label{Ez29}
\end{eqnarray}
where in the last step one uses the second line of (\ref{Ez11}). From here,
introducing the notations$X=\frac{t^2}{|a_{1,1}|^2}$, and $Y=|a_{2,2}|^2$ ,
the following system of linear equations is found
\begin{eqnarray}
&&\epsilon_2=X+Y,
\nonumber\\
&&Y=\frac{t^2_h}{\epsilon_2} +\frac{t_h X}{\epsilon_2},
\label{Ez30}
\end{eqnarray}
from where
\begin{eqnarray}
&&X=\frac{t^2}{|a_{1,1}|^2} = \epsilon_2-t_h, \: => \:
a_{1,1}= \frac{|t|e^{i\theta_1}}{\sqrt{\epsilon_2-t_h}}, \: \epsilon_2 > t_h > 0,
\nonumber\\
&&Y=|a_{2,2}|^2 =t_h, \: => \: a_{2,2}=\sqrt{t_h} e^{i\theta_2}.
\label{Ez31}
\end{eqnarray}
Here, $\theta_1,\theta_2$ are arbitrary phases. Now, from the first two
equations of (\ref{Ez11}), one finds
\begin{eqnarray}
a_{2,5}=-\frac{\epsilon_2-t_h}{\sqrt{t_h}}e^{i\theta_2}, \quad
a_{3,4}=\frac{|t|}{\sqrt{\epsilon_2-t_h}}e^{i\theta_3}
\label{Ez32}
\end{eqnarray}
where again, $\theta_3$ is an arbitrary phase. One underlines, that after all
these steps, from (\ref{Ez10}) only the third equation remains non-analyzed.

Now, using the first line of (\ref{Ez6}), from (\ref{Ez9}) one finds
\begin{eqnarray}
&&a_{1,2}=a_{1,5}=sign(t)\sqrt{\epsilon_2-t_h} \: e^{i\theta_1},
a_{3,3}=a_{3,5}=sign(t)\sqrt{\epsilon_2-t_h} \: e^{i\theta_3}, 
\nonumber\\
&&a_{2,2}=a_{2,3}=\sqrt{t_h} \: e^{i\theta_2}, \:
a_{4,6}=\sqrt{\epsilon_4} \: e^{i\theta_4}, \: 
a_{4,5}=\frac{t_f}{\sqrt{\epsilon_4}} \: e^{i\theta_4},
\label{Ez33}
\end{eqnarray}
where $\theta_4$ is an arbitrary phase.

\subsubsection*{3.c) The results relating the b coefficients}

Using (\ref{Ez16}), the first two equalities of (\ref{Ez19}) provide
\begin{eqnarray}
b_{1,1}=sign(t) \frac{t_{2,1}^{\uparrow,\downarrow}}{\sqrt{\epsilon_2-t_h}}e^{i\theta_1},
\: b'_{1,1}=e^{-i\phi_{\alpha}} b_{1,1}, \:
b_{3,4}= sign(t) \frac{t_{5,4}^{\uparrow,\downarrow}}{\sqrt{\epsilon_2-t_h}}
e^{i\theta_3}, \: b'_{3,4}= e^{-i\phi_{\alpha}} b_{3,4},
\label{Ez34}
\end{eqnarray}
One further notes, that from (\ref{Ez19}), only one equality remains to be
analyzed, namely the third one.

\subsubsection*{3.d) The remaining equations}

After these steps we underline, that till now only seven equations have not been
used, namely, the following ones
\begin{eqnarray}
&&t_c = a^*_{5,7}a_{5,4}+ b^*_{5,7}b_{5,4},
\nonumber\\
&&t^{\uparrow,\downarrow}_c={b}^*_{5,7}a_{5,4}e^{i\phi_{\alpha}}+a^*_{5,7}b_{5,4},
\nonumber\\
&&(a^*_{5,7}b_{5,7}+b^*_{5,7}a_{5,7}e^{i\phi_{\alpha}})=-
(a^*_{1,1}b_{1,1}+b^*_{1,1}a_{1,1}e^{i\phi_{\alpha}}),
\nonumber\\
&&(a^*_{5,4}b_{5,4}+b^*_{5,4}a_{5,4}e^{i\phi_{\alpha}})=-(a^*_{3,4}b_{3,4}+b^*_{3,4}a_{3,4}
e^{i\phi_{\alpha}}),
\nonumber\\
&&\epsilon_3=\epsilon^{\uparrow,\uparrow}_5=\epsilon^{\downarrow,\downarrow}_5=|a_{4,5}|^2
+|a_{1,5}|^2+|a_{2,5}|^2+|a_{3,5}|^2,
\nonumber\\
&&\epsilon_1=\epsilon^{\downarrow,\downarrow}_1=|a_{1,1}|^2
+|a_{5,7}|^2+|b_{1,1}|^2+|b_{5,7}|^2,
\nonumber\\
&&\epsilon_1=\epsilon^{\downarrow,\downarrow}_4=|a_{3,4}|^2
+|a_{5,4}|^2+|b_{3,4}|^2+|b_{5,4}|^2.
\label{Ez35}
\end{eqnarray}
Introducing here the deduced results one finds the following system of equations
\begin{eqnarray}
&&t_c = a^*_{5,7}a_{5,4}+ b^*_{5,7}b_{5,4},
\nonumber\\
&&t^{\uparrow,\downarrow}_c={b}^*_{5,7}a_{5,4}e^{i\phi_{\alpha}}+a^*_{5,7}b_{5,4},
\nonumber\\
&&(a^*_{5,7}b_{5,7}+b^*_{5,7}a_{5,7}e^{i\phi_{\alpha}})=-\frac{t}{\epsilon_2-t_h}(
t^{\uparrow,\downarrow}_{2,1} + {t^{\uparrow,\downarrow}_{2,1}}^*e^{i\phi_{\alpha}}),
\nonumber\\
&&(a^*_{5,4}b_{5,4}+b^*_{5,4}a_{5,4}e^{i\phi_{\alpha}})=-\frac{t}{\epsilon_2-t_h}(
t^{\uparrow,\downarrow}_{5,4} + {t^{\uparrow,\downarrow}_{5,4}}^*e^{i\phi_{\alpha}}),
\nonumber\\
&&\epsilon_3= \frac{t_f^2}{\epsilon_4} + \frac{\epsilon^2_2-t_h^2}{t_h},
\nonumber\\
&&\epsilon_1 - \frac{t^2+{t^{\uparrow,\downarrow}_{2,1}}^2}{\epsilon_2-t_h}=
|a_{5,7}|^2+|b_{5,7}|^2,
\nonumber\\
&&\epsilon_1 - \frac{t^2+{t^{\uparrow,\downarrow}_{5,4}}^2}{\epsilon_2-t_h}=
|a_{5,4}|^2+|b_{5,4}|^2.
\label{Ez36}
\end{eqnarray}
In our case $t^{\uparrow,\downarrow}$ are real parameters, and $e^{i \phi_\alpha} = -1$
holds. Hence the third and fourth equations of (\ref{Ez36}) provide
\begin{eqnarray}
&&a^*_{5,7} b_{5,7} = a_{5,7} b^*_{5,7} ,
\nonumber\\
&&a^*_{5,4}{b}_{5,4} = a_{5,4}b^*_{5,4}.
\label{Ez37}
\end{eqnarray}
But this means that $a_{5,7} , b_{5,7}$ have equal phases, and similarly,
$a_{5,4} , b_{5,4}$ have equal phases. I.e.
\begin{eqnarray}
&&a_{5,7} = |a_{5,7}| e^{i \chi} , \quad b_{5,7} = |b_{5,7}| e^{i \chi} ,
\nonumber\\
&&a_{5,4} = |a_{5,4}| e^{i \eta} , \quad b_{5,4} = |b_{5,4}| e^{i \eta} ,
\label{Ez38}
\end{eqnarray}
Introducing now (\ref{Ez38}) in the first equality of (\ref{Ez36}),
one finds
\begin{eqnarray}
&&t_c = |a_{5,7}| e^{-i \chi} |a_{5,4}| e^{i \eta} + |b_{5,7}| e^{-i \chi} 
|b_{5,4}| e^{i \eta},
\nonumber\\
&&\chi = \eta \pm 2 n \pi, n \in N.
\label{Ez39}
\end{eqnarray}
Now one introduces two notations, namely $\gamma_1$ and $\gamma_2$ defined as
\begin{eqnarray}
&&\gamma_1 = \epsilon_1 - \frac{t^2+{t^{\uparrow,\downarrow}_{2,1}}^2}{\epsilon_2-t_h},
\nonumber\\
&&\gamma_2 = \epsilon_1 - \frac{t^2+{t^{\uparrow,\downarrow}_{5,4}}^2}{\epsilon_2-t_h}
\label{Ez40}
\end{eqnarray}
In this manner, the first, and last two equations from (\ref{Ez36}) build
up a closed system of equations
\begin{eqnarray}
&&t_c = |a_{5,7}|  |a_{5,4}|  + |b_{5,7}| |b_{5,4}| ,
\nonumber\\
&&t^{\uparrow,\downarrow}_c=  |a_{5,7}||b_{5,4}| - |{b}_{5,7}||a_{5,4}| ,
\nonumber\\
&&\gamma_1 =
|a_{5,7}|^2+|b_{5,7}|^2,
\nonumber\\
&&\gamma_2 =
|a_{5,4}|^2+|b_{5,4}|^2.
\label{Ez41}
\end{eqnarray}
In what will follows, we must solve the system of equations (\ref{Ez41})
In order to do this job, we must observe, that (\ref{Ez41}) not contains
independent equations because it uses a relation in between the Hamiltonian
parameters. Indeed, taking the square of the first two equations and adding
them, one finds
\begin{eqnarray}
&&|a_{5,7}|^2|a_{5,4}|^2 +|b_{5,7}|^2|b_{5,4}|^2 + |a_{5,7}|^2|b_{5,4}|^2 +
|b_{5,7}|^2|a_{5,4}|^2 =t^2_c +{t^{\uparrow,\downarrow}_c}^2, \: \to
\nonumber\\
&&|a_{5,7}|^2(|a_{5,4}|^2+|b_{5,4}|^2)+|b_{5,7}|^2|(|a_{5,4}|^2+|b_{5,4}|^2)=
t^2_c +{t^{\uparrow,\downarrow}_c}^2, \: \to
\nonumber\\
&&\gamma_1 \gamma_2 =t^2_c +{t^{\uparrow,\downarrow}_c}^2.
\label{Ez42}
\end{eqnarray}
From the second equation of (\ref{Ez41})
$|b_{5,4}| = \frac{t_c^{\uparrow \downarrow} + |b_{5,7}| |a_{5,4}|  }{|a_{5,7}|}$,
and this introduced in the first equality, allows to obtain the expression of
$|b_{5,7}|$ in function of the $|a|$ coefficients. Using also the $\gamma_2$
expression, i.e. the fourth equation of (\ref{Ez41}), one obtains
\begin{eqnarray}
&&|b_{5,7}| = \frac{t_c |a_{5,7}| - \gamma_2 |a_{5,4}| }{t_c^{\uparrow \downarrow} }.
\label{Ez43}
\end{eqnarray}
Similarly, from the second equation of (\ref{Ez41}) one has 
$|b_{5,7}| = \frac{-t_c^{\uparrow \downarrow} + |b_{5,4}| |a_{5,7}|  }{|a_{5,4}|}$.
This introduced in the first equation one can express $|b_{5,7}|$ in function
of $|a|$ parameters, and using the $\gamma_1$ expression from the third
equation one finds
\begin{eqnarray}
&&|b_{5,4}| = \frac{\gamma_1|a_{5,7}| - t_c |a_{5,4}| }{t_c^{\uparrow \downarrow} }.
\label{Ez44}
\end{eqnarray}
In this moment the last two equations from (\ref{Ez41}) must be analyzed.
Introducing here all previously deduced results, and introducing the notations
$x=|a_{5,4}|$, $y=|a_{5,7}|$, the following system of equations are obtained 
\begin{eqnarray}
&&\gamma_1 =
x^2+\bigg(\frac{\gamma_1 y - t_c x}{t_c^{\uparrow,\downarrow}}\bigg)^2,
\nonumber\\
&&\gamma_2 =
y^2+\bigg(\frac{t_c y - \gamma_2 x}{t_c^{\uparrow,\downarrow}}\bigg)^2.
\label{Ez45}
\end{eqnarray}
Now it can be observed that the here obtained two equations, via
$y \to \sqrt{\gamma_2/\gamma_1} \: x, \: x \to \sqrt{\gamma_1/\gamma_2} \: y$,
transform in themselves, hence
\begin{eqnarray}
|a_{5,7}|=y= \sqrt{\gamma_2/\gamma_1} \: x, \: x=|a_{5,4}|
\label{Ez46}
\end{eqnarray}
Using (\ref{Ez46})), the relation (\ref{Ez44}) can be written as
\begin{eqnarray}
\gamma_1 = x^2 + \frac{x^2}{{t^{\uparrow,\downarrow}_c}^2} [ \gamma_1
\sqrt{\frac{\gamma_2}{\gamma_1}}-t_c]^2.
\label{Ez47}
\end{eqnarray}
Based on (\ref{Ez47}) and using (\ref{Ez42}), one finds
\begin{eqnarray}
&&|a_{5,4}|=x =\frac{\sqrt{\gamma_1}}{\sqrt{1+\frac{(\sqrt{\gamma_1\gamma_2}
-t_c)^2}{{t^{\uparrow,\downarrow}_c}^2}}}, \quad
|a_{5,7}|=y =\frac{\sqrt{\gamma_2}}{\sqrt{1+\frac{(\sqrt{\gamma_1\gamma_2}-t_c)^2
}{{t^{\uparrow,\downarrow}_c}^2}}},
\nonumber\\
&&\gamma_1\gamma_2=t^2_c +{t^{\uparrow,\downarrow}_c}^2.
\label{Ez48}
\end{eqnarray}

\subsection{Summary of the obtained results}

\subsubsection{The deduced block operator parameters.}

The analyzed matching equations provided the following results: 
\begin{eqnarray}
&&a_{1,1}= \frac{|t|e^{i\theta_1}}{\sqrt{\epsilon_2-t_h}}, \:
a_{1,2}=a_{1,5}=sign(t)\sqrt{\epsilon_2-t_h} \: e^{i\theta_1},
\nonumber\\
&&a_{2,5}=-\frac{\epsilon_2-t_h}{\sqrt{t_h}}e^{i\theta_2}, \:
a_{2,2}=a_{2,3}=\sqrt{t_h} \: e^{i\theta_2}, 
\nonumber\\
&&a_{3,4}=\frac{|t|}{\sqrt{\epsilon_2-t_h}}e^{i\theta_3}, \:
a_{3,3}=a_{3,5}=sign(t)\sqrt{\epsilon_2-t_h} \: e^{i\theta_3},
\nonumber\\
&&a_{4,6}=\sqrt{\epsilon_4} \: e^{i\theta_4}, \: 
a_{4,5}=\frac{t_f}{\sqrt{\epsilon_4}} \: e^{i\theta_4},
\nonumber\\
&&a_{5,4}=\frac{\sqrt{\gamma_1}}{\sqrt{1+\frac{(\sqrt{\gamma_1\gamma_2}
-t_c)^2}{{t^{\uparrow,\downarrow}_c}^2}}} e^ {i\theta_5}, \:
a_{5,7}=\frac{\sqrt{\gamma_2}}{\sqrt{1+\frac{(\sqrt{\gamma_1\gamma_2}-t_c)^2
}{{t^{\uparrow,\downarrow}_c}^2}}} e^{i\theta_5},
\nonumber\\
&&b_{1,1}=sign(t) \frac{t_{2,1}^{\uparrow,\downarrow}}{\sqrt{\epsilon_2-t_h}}e^{i\theta_1},
\: b'_{1,1}=-sign(t) \frac{t_{2,1}^{\uparrow,\downarrow}}{\sqrt{\epsilon_2-t_h}}
e^{i\theta_1},
\nonumber\\
&&b_{3,4}= sign(t) \frac{t_{5,4}^{\uparrow,\downarrow}}{\sqrt{\epsilon_2-t_h}}
e^{i\theta_3}, \: b'_{3,4}= -sign(t) \frac{t_{5,4}^{\uparrow,\downarrow}}{\sqrt{
\epsilon_2-t_h}}e^{i\theta_3},
\nonumber\\
&&b_{5,4}=\frac{\gamma_1\sqrt{\gamma_2}-t_c \sqrt{\gamma_1}}{t^{\uparrow,\downarrow}_c
\sqrt{1+\frac{(\sqrt{\gamma_1\gamma_2}-t_c)^2}{{t^{\uparrow,\downarrow}_c}^2}}}
e^{i\theta_5}, \:
b'_{5,4}=-\frac{\gamma_1\sqrt{\gamma_2}-t_c \sqrt{\gamma_1}}{t^{\uparrow,\downarrow}_c
\sqrt{1+\frac{(\sqrt{\gamma_1\gamma_2}-t_c)^2}{{t^{\uparrow,\downarrow}_c}^2}}}
e^{i\theta_5}, 
\nonumber\\
&&b_{5,7}=\frac{t_c\sqrt{\gamma_2}-\gamma_2 \sqrt{\gamma_1}}{t^{\uparrow,\downarrow}_c
\sqrt{1+\frac{(\sqrt{\gamma_1\gamma_2}-t_c)^2}{{t^{\uparrow,\downarrow}_c}^2}}}
e^{i\theta_5}, \:
b'_{5,7}=-\frac{t_c\sqrt{\gamma_2}-\gamma_2 \sqrt{\gamma_1}}{t^{\uparrow,\downarrow}_c
\sqrt{1+\frac{(\sqrt{\gamma_1\gamma_2}-t_c)^2}{{t^{\uparrow,\downarrow}_c}^2}}}
e^{i\theta_5}.
\label{Ez48}
\end{eqnarray}
The deduced solution is placed in the following parameter domain
\begin{eqnarray}
&&\epsilon_1 > 0, \: \epsilon_2 > 0, \: \epsilon_3 > 0, \: \epsilon_4 > 0, \:
 t_h > 0, \: 
\epsilon_2-t_h > 0,
\nonumber\\
&&\gamma_1=\epsilon_1 - \frac{t^2+{t^{\uparrow,\downarrow}_{2,1}}^2}{\epsilon_2-t_h} 
> 0, \:
\gamma_2=\epsilon_1 - \frac{t^2+{t^{\uparrow,\downarrow}_{5,4}}^2}{\epsilon_2-t_h} > 0,
\nonumber\\
&&\epsilon_3= \frac{t_f^2}{\epsilon_4} + \frac{\epsilon^2_2-t_h^2}{t_h}, \:
\gamma_1 \gamma_2 =t^2_c +{t^{\uparrow,\downarrow}_c}^2.
\label{Ez49}
\end{eqnarray}
I note that from symmetry considerations one has $t^{\uparrow,\downarrow}_{2,1}=
t^{\uparrow,\downarrow}_{5,4}=t^{\uparrow,\downarrow}$, hence $\gamma_1=\gamma_2=\gamma$
holds
\begin{eqnarray}
\gamma_1=\gamma_2=\gamma.
\label{Ez50}
\end{eqnarray}

\subsubsection{The check of the solutions of the matching equations}

Using this matching equation solution, one exemplifies how the obtained
results can be checked.
First we check the obtained results for the $z=5$ block. One starts
with the equation written for the $t_c$ hopping element.
\begin{eqnarray}
&&t_c=a^*_{5,7}a_{5,4}+b^*_{5,7}b_{5,4}=\frac{\sqrt{\gamma_1\gamma_2}}{1+\frac{(
\sqrt{\gamma_1
\gamma_2}-t_c)^2}{{t^{\uparrow,\downarrow}_c}^2}} + \frac{(\gamma_1\sqrt{\gamma_2}-
t_c\sqrt{\gamma_1})(t_c\sqrt{\gamma_2}-\gamma_2 \sqrt{\gamma_1})}{
{t^{\uparrow,\downarrow}_c}^2(1+
\frac{(\sqrt{\gamma_1\gamma_2}-t_c)^2}{{t^{\uparrow,\downarrow}_c}^2})} =
\nonumber\\
&&\frac{{t^{\uparrow,\downarrow}_c}^2\sqrt{\gamma_1\gamma_2}+(2t_c\gamma_1\gamma_2-
t^2_c\sqrt{\gamma_1\gamma_2}-\gamma_1\gamma_2\sqrt{\gamma_1\gamma_2})}{
{t^{\uparrow,\downarrow}_c}^2+(\sqrt{\gamma_1\gamma_2}-t_c)^2}=
\nonumber\\
&&\frac{(\gamma_1\gamma_2-t_c^2)
\sqrt{\gamma_1\gamma_2}+(2t_c\gamma_1\gamma_2-
t^2_c\sqrt{\gamma_1\gamma_2}-\gamma_1\gamma_2\sqrt{\gamma_1\gamma_2})}{
(\gamma_1\gamma_2-t^2_c)+(\sqrt{\gamma_1\gamma_2}-t_c)^2}=
\nonumber\\
&&\frac{t_c(2\gamma_1\gamma_2-2t_c\sqrt{\gamma_1\gamma_2})}{
(2\gamma_1\gamma_2-2t_c\sqrt{\gamma_1\gamma_2})}=t_c
\label{Ez51}
\end{eqnarray}
After this step the equation of $t^{\uparrow,\downarrow}_c$ follows:
\begin{eqnarray}
&&t^{\uparrow,\downarrow}_c={b'}^*_{5,7}a_{5,4}+a^*_{5,7}b_{5,4}=
-\frac{(t_c\sqrt{\gamma_2}-\gamma_2\sqrt{\gamma_1})
\sqrt{\gamma_1}}{t^{\uparrow,\downarrow}_c(1+
\frac{(\sqrt{\gamma_1\gamma_2}-t_c)^2}{{t^{\uparrow,\downarrow}_c}^2})} +
\frac{(\gamma_1\sqrt{\gamma_2}-t_c\sqrt{\gamma_1})
\sqrt{\gamma_2}}{t^{\uparrow,\downarrow}_c(1+
\frac{(\sqrt{\gamma_1\gamma_2}-t_c)^2}{{t^{\uparrow,\downarrow}_c}^2})} =
\nonumber\\
&& t^{\uparrow,\downarrow}_c \frac{(-t_c\sqrt{\gamma_1\gamma_2}+\gamma_1\gamma_2) +
(\gamma_1\gamma_2-t_c\sqrt{\gamma_1\gamma_2})}{{t^{\uparrow,\downarrow}_c}^2 +
(\sqrt{\gamma_1\gamma_2}-t_c)^2}= t^{\uparrow,\downarrow}_c \frac{2\gamma_1\gamma_2-
2t_c\sqrt{\gamma_1\gamma_2}}{(\gamma_1\gamma_2-t^2_c) +(\sqrt{\gamma_1\gamma_2}-
t_c)^2}=
\nonumber\\
&&t^{\uparrow,\downarrow}_c \frac{2\gamma_1\gamma_2-2t_c\sqrt{\gamma_1\gamma_2}}{
2\gamma_1\gamma_2-2t_c\sqrt{\gamma_1\gamma_2}}=t^{\uparrow,\downarrow}_c.
\label{Ez52}
\end{eqnarray}
After this step one checks the two equations obtained for $\epsilon_1$,
(see also (\ref{Ez51}))
\begin{eqnarray}
&&\gamma=\gamma_1=|a_{5,7}|^2+|b_{5,7}|^2=\frac{\gamma_2}{1+
\frac{(\sqrt{\gamma_1\gamma_2}-t_c)^2}{{t^{\uparrow,\downarrow}_c}^2}} +
\frac{(t_c\sqrt{\gamma_2}-\gamma_2\sqrt{\gamma_1})^2}{
{t^{\uparrow,\downarrow}_c}^2(1+
\frac{(\sqrt{\gamma_1\gamma_2}-t_c)^2}{{t^{\uparrow,\downarrow}_c}^2})}=
\nonumber\\
&&\gamma_2\frac{{t^{\uparrow,\downarrow}_c}^2+(t^2_c+\gamma_2\gamma_1-2t_c \sqrt{
\gamma_2\gamma_1})}{{t^{\uparrow,\downarrow}_c}^2+(\sqrt{\gamma_2\gamma_1}-t_c)^2}=
\gamma_2=\gamma.
\label{Ez53}
\end{eqnarray}
Similar result is obtained from the second equality written for $\epsilon_1$
(see the last equality of (\ref{Ez41}))

And finally, the  $\epsilon^{\uparrow,\downarrow}_1$ equation (see (\ref{Ez7}))
\begin{eqnarray}
&&\epsilon^{\uparrow,\downarrow}_1=a^*_{1,1}b_{1,1}+a^*_{5,7}b_{5,7}+{b'}^*_{1,1}a_{1,1}+
{b'}^*_{5,7}a_{5,7}=
\nonumber\\
&&|a_{1,1}||b_{1,1}|-|a_{1,1}||b_{1,1}|+|a_{5,7}||b_{5,7}|-
|a_{5,7}||b_{5,7}|=0.
\label{Ez54}
\end{eqnarray}

The other matching conditions are trivially satisfied.

\subsubsection{Final results relating the block operator parameters}

If one uses all the deduced results, including the symmetry considerations,
the block operator parameters can be
expressed in final form as follows:
\begin{eqnarray}
&&a_{1,1}= \frac{|t|e^{i\theta_1}}{\sqrt{\epsilon_2-t_h}}, \:
a_{1,2}=a_{1,5}=sign(t)\sqrt{\epsilon_2-t_h} \: e^{i\theta_1},
\nonumber\\
&&a_{2,5}=-\frac{\epsilon_2-t_h}{\sqrt{t_h}}e^{i\theta_2}, \:
a_{2,2}=a_{2,3}=\sqrt{t_h} \: e^{i\theta_2}, 
\nonumber\\
&&a_{3,4}=\frac{|t|}{\sqrt{\epsilon_2-t_h}}e^{i\theta_3}, \:
a_{3,3}=a_{3,5}=sign(t)\sqrt{\epsilon_2-t_h} \: e^{i\theta_3},
\nonumber\\
&&a_{4,6}=\sqrt{\epsilon_4} \: e^{i\theta_4}, \: 
a_{4,5}=\frac{t_f}{\sqrt{\epsilon_4}} \: e^{i\theta_4},
\nonumber\\
&&a_{5,4}=a_{5,7}=\sqrt{\frac{\gamma+t_c}{2}} e^ {i\theta_5},
\nonumber\\
&&b_{1,1}=sign(t) \frac{t^{\uparrow,\downarrow}}{\sqrt{\epsilon_2-t_h}}e^{i\theta_1},
\: b'_{1,1}=-sign(t) \frac{t^{\uparrow,\downarrow}}{\sqrt{\epsilon_2-t_h}}
e^{i\theta_1},
\nonumber\\
&&b_{3,4}= sign(t) \frac{t^{\uparrow,\downarrow}}{\sqrt{\epsilon_2-t_h}}
e^{i\theta_3}, \: b'_{3,4}= -sign(t) \frac{t^{\uparrow,\downarrow}}{\sqrt{
\epsilon_2-t_h}}e^{i\theta_3},
\nonumber\\
&&b_{5,4}=\frac{\gamma-t_c}{t^{\uparrow,\downarrow}_c}\sqrt{\frac{\gamma+t_c}{2}} 
e^ {i\theta_5}, \:
b'_{5,4}=-\frac{\gamma-t_c}{t^{\uparrow,\downarrow}_c}\sqrt{\frac{\gamma+t_c}{2}} 
e^ {i\theta_5},
\nonumber\\
&&b_{5,7}=\frac{t_c-\gamma}{t^{\uparrow,\downarrow}_c}\sqrt{\frac{\gamma+t_c}{2}} 
e^ {i\theta_5}, \:
b'_{5,7}=-\frac{t_c-\gamma}{t^{\uparrow,\downarrow}_c}\sqrt{\frac{\gamma+t_c}{2}} 
e^ {i\theta_5}.
\label{Ez55}
\end{eqnarray}
The conditions under which the deduced solution hold (i.e. the parameter space
domain where the solution is valid) can be given as follows
\begin{eqnarray}
&&\epsilon_1 > 0, \: \epsilon_4 > 0,
\: t_h > 0, \: 
\epsilon_2-t_h > 0,
\nonumber\\
&&\gamma=\epsilon_1 - \frac{t^2+{t^{\uparrow,\downarrow}}^2}{\epsilon_2-t_h}, 
\gamma+t_c > 0,
\nonumber\\
&&\epsilon_3= \frac{t_f^2}{\epsilon_4} + \frac{\epsilon^2_2-t_h^2}{t_h}, \:
\gamma^2=t^2_c +{t^{\uparrow,\downarrow}_c}^2.
\label{Ez56}
\end{eqnarray}

\section{Solution of the matching equations at $B \ne 0$}

We start by (\ref{E15}) from where all b coefficients can be written in terms
of the a coefficients
\begin{eqnarray}
&& {b'}^*_{1,1}= - \frac{\lambda e^{i\phi_3}}{a_{1,5,d}}, \:
b^*_{1,1}=  \frac{\lambda e^{i\phi_3}}{a_{1,5,u}}, \:
b^*_{1,1}=  \frac{\lambda e^{i\phi_2}}{a^*_{1,2,u}}, \:
b'_{1,1}= - \frac{\lambda e^{i\phi_2}}{a^*_{1,2,d}}, \:
b_{3,4}= - \frac{\lambda e^{i\phi_3}}{a^*_{3,5,u}},
\label{C1}\\
&& {b'}_{3,4}=  \frac{\lambda e^{i\phi_3}}{a^*_{3,5,d}}, \:
{b'}^*_{3,4}=  \frac{\lambda e^{i\phi_2}}{a_{3,3,d}}, \:
b^*_{3,4}= - \frac{\lambda e^{i\phi_2}}{a_{3,3,u}}, \:
{b'}^*_{5,7}= \frac{\lambda_c-a^*_{5,7,u}b_{5,4}}{a_{5,4,d}}, \:
b'_{5,4}= \frac{-\lambda_c -a_{5,4,u}b^*_{5,7}}{a^*_{5,7,d}},
\nonumber
\end{eqnarray}
so we can concentrate (at least up to the study of the last two bond operator
parameters from (\ref{Ez2})) to the $a_{i,j,\alpha}$ parameters.

Now, we write the first 12 equations from (\ref{Ez14}) in the following form
\begin{eqnarray}
&&a_{1,5,u}=\frac{te^{i\phi_3}}{a^*_{1,1,u}}, \: a^*_{1,2,u}=\frac{te^{i\phi_2}}{
a_{1,1,u}}, \: a_{1,5,d}=\frac{te^{i\phi_3}}{a^*_{1,1,d}}, \:
a^*_{1,2,d}=\frac{te^{i\phi_2}}{a^*_{1,1,d}},
\nonumber\\
&&a_{3,3,u}=\frac{te^{i\phi_2}}{a^*_{3,4,u}}, \: a^*_{3,5,u}=\frac{te^{i\phi_3}}{
a_{3,4,u}}, \: a_{3,3,d}=\frac{te^{i\phi_2}}{a^*_{3,4,d}}, \:
a^*_{3,5,d}=\frac{te^{i\phi_3}}{a^*_{3,4,d}},
\nonumber\\
&&a^*_{2,3,u}=\frac{t_he^{i\phi_1}}{a_{2,2,u}}, \: a^*_{2,3,d}=\frac{t_he^{i\phi_1}}{
a_{2,2,d}}, \: a_{4,5,u}=\frac{t_f}{a^*_{4,6,u}}, \:
a_{4,5,d}=\frac{t_f}{a^*_{4,6,d}}.
\label{C2}
\end{eqnarray}
From the last two equations of (\ref{C1}) and the first two equations of
(\ref{E16}) we automatically obtains
\begin{eqnarray}
a_{4,5,u}=\frac{t_f e^{i\chi_{4,u}}}{\sqrt{\epsilon_6-h}}, \:
a_{4,5,d}=\frac{t_f e^{i\chi_{4,d}}}{\sqrt{\epsilon_6+h}}, \:
a_{4,6,u}= e^{i\chi_{4,u}} \sqrt{\epsilon_6-h}, \:
a_{4,6,d}= e^{i\chi_{4,d}} \sqrt{\epsilon_6+h}, 
\label{C3}
\end{eqnarray}
where $\chi_{4,\nu}$, $\nu=u,d$ are arbitrary phases. So the $z=4$ block
operator parameters have been deduced. One what will follows one concentrates
on the $z=1,2,3$ block operators. We present in details the $\nu=u$ case, and
underline that the deduction of the parameters in the $\nu=d$ case is done
perfectly similarly.

One starts with the second and fourth equality from the bottom of (\ref{E14})
obtaining
\begin{eqnarray}
&& a_{2,5,u} =- \frac{a^*_{1,2,u}a_{1,5,u}}{a^*_{2,2,u}}= - \frac{1}{a^*_{2,2,u}}
\frac{t^2 e^{i(\phi_2+\phi_3)}}{|a_{1,1,u}|^2},
\nonumber\\
&& a^*_{2,5,u} =- \frac{a^*_{3,5,u}a_{3,3,u}}{a_{2,3,u}}= - \frac{1}{a_{2,3,u}}
\frac{t^2 e^{i(\phi_2+\phi_3)}}{|a_{3,4,u}|^2} = -\frac{a^*_{2,2,u}}{t_h e^{-i\phi_1}}
\frac{t^2 e^{i(\phi_2+\phi_3)}}{|a_{3,4,u}|^2},
\nonumber\\
&&|a_{2,5,u}|^2 = \frac{t^4 e^{i(\phi_1+2\phi_2 + 2\phi_3)}}{t_h |a_{1,1,u}|^2
  |a_{3,4,u}|^2}=
\frac{t^4}{T_h |a_{1,1,u}|^2|a_{3,4,u}|^2}, \: T_h = \frac{|a_{2,2,u}|^2|a_{1,1,u}|^2}{
  |a_{3,4,u}|^2}.
\label{C4} 
\end{eqnarray}
where $T_h = t_h e^{-i\phi}, \phi = \phi_1+2\phi_2+2\phi_3 > 0$ must hold.
Here, in the first two rows we have used (\ref{C3}) as well, while in the last
row, we multiplied the first two rows. I also mention, that from the second
equality of the first row from (\ref{C4}) one has $|a_{2,5,u}|^2=t^4/(|a_{2,2,u}|^2
|a_{1,1,u}|^4)$, from where the last equality of (\ref{C4}) follows.

Now we use the 5th and 7th equation from (\ref{E16}), obtaining (in the last
step, the last equality of (\ref{C4}) is used)
\begin{eqnarray}
&&\epsilon_2-h=|a_{1,2,u}|^2 + |a_{2,2,u}|^2=\frac{t^2}{|a_{1,1,u}|^2} + |a_{2,2,u}|^2,
\nonumber\\
&&\epsilon_3-h=|a_{2,3,u}|^2 + |a_{3,3,u}|^2=\frac{t_h^2}{|a_{2,2,u}|^2} +
\frac{t^2}{|a_{3,4,u}|^2}= \frac{t_h^2}{|a_{2,2,u}|^2} + \frac{t^2 T_h}{
|a_{2,2,u}|^2 |a_{1,1,u}|^2}.
\label{C5}
\end{eqnarray}
From (\ref{C5}) now $a_{1,1,u}, a_{2,2,u}$ can be expressed
\begin{eqnarray}
a_{1,1,u}= \sqrt{\frac{t^2(\epsilon_3-h+T_h)}{(\epsilon_2-h)(\epsilon_3-h)-t_h^2)}}
e^{i\chi_{1,u}}, \quad a_{2,2,u}= \sqrt{\frac{T_h(\epsilon_2-h)+t_h^2}{\epsilon_3-
h+T_h}}e^{i\chi_{2,u}}.
\label{C6}
\end{eqnarray}
Based on the last two equations of (\ref{C4}) now $a_{3,4,u}$ (from the $T_h$
expression), then $a_{2,5,u}$ (from the first line of (\ref{C4})) can be
expressed
\begin{eqnarray}
a_{3,4,u} = \sqrt{\frac{t^2[T_h(\epsilon_2-h)+t_h^2]}{T_h[(\epsilon_2-h)(
\epsilon_3-h)-t_h^2]}} e^{i\chi_{3,u}}, \: a_{2,5,u}= - e^{i(\phi_2+\phi_3)}
\frac{ e^{i\chi_{2,u}} [(\epsilon_2-h)(\epsilon_3-h)-t_h^2]}{\sqrt{T_h(\epsilon_2-h)
+t_h^2}\sqrt{\epsilon_3-h+T_h}}.
\label{C7}
\end{eqnarray}  
At this moment $a_{1,2,u},a_{1,5,u},a_{2,3,u},a_{3,4,u},a_{3,5,u}$ can be deduced from
(\ref{C2}), while $b_{1,1},b_{3,4}$ are directly obtained from (\ref{C1}).

Starting from the 6th and 8th equalities of (\ref{E16}),and the first and third
equality from the bottom of (\ref{E14}) a similar deduction provides the
coefficients with index d, for z=1,2,3.

In the case of the block eoperator coefficients connected to the bond 4-7,
namely the z=5 block operators, the procedure is more complicated, and
different. The used equations in this case are the last six equalities from
(\ref{E16}), the last two equalities of (\ref{E15}) and the two equations
containing $t_c$ from (\ref{E14}). These equations can be written as
\begin{eqnarray}
&&I_{1,u}=|a_{5,7,u}|^2 + |b'_{5,7}|^2, \quad I_{1,d}=|a_{5,7,d}|^2 + |b_{5,7}|^2,
\nonumber\\
&&I_{2,u}=|a_{5,4,u}|^2 + |b'_{5,4}|^2, \quad I_{2,d}=|a_{5,4,d}|^2 + |b_{5,4}|^2,
\nonumber\\
&&V_1 =a^*_{5,7,u}b_{5,7} + {b'}^*_{5,7}a_{5,7,d}, \quad 
V_2 =a^*_{5,4,u}b_{5,4} + {b'}^*_{5,4}a_{5,4,d},
\nonumber\\
&&\lambda_c =  {b'}^*_{5,7}a_{5,4,d} + a^*_{5,7,u}b_{5,4}, \quad
-\lambda_c =  b^*_{5,7}a_{5,4,u} + a^*_{5,7,d}b'_{5,4},
\nonumber\\
&&t_c= a^*_{5,7,u}a_{5,4,u} + {b'}^*_{5,7}b'_{5,4}, \quad
t_c= a^*_{5,7,d}a_{5,4,d} + b^*_{5,7}b_{5,4},
\label{C8}
\end{eqnarray} 
where $I_{1,u},I_{2,u},I_{1,d},I_{2,d},V_1,V_2$ have been defined in
(\ref{E21}), and based on the results presented above (i.e. (\ref{C1})-
(\ref{C7})), are known quantities. I exemplify the deduction procedure
leading to the $W_1$ expression from (\ref{E22}) and the first equality
from (\ref{E23}), namely $b'_{5,7}=W_1 a_{5,4,d}$.

One starts from the third and fourth row of (\ref{C8}) which provide
\begin{eqnarray}
&& b_{5,4}=\frac{V_2 -{b'}^*_{5,4}a_{5,4,d}}{a^*_{5,4,u}} =
\frac{\lambda_c - {b'}^*_{5,7}a_{5,4,d}}{a^*_{5,7,u}}
\nonumber\\
&& b_{5,7}=\frac{V_1 -{b'}^*_{5,7}a_{5,7,d}}{a^*_{5,7,u}} =
\frac{-\lambda_c - {b'}^*_{5,4}a_{5,7,d}}{a^*_{5,4,u}}
\label{C9}  
\end{eqnarray}
from where one finds
\begin{eqnarray}
&& \frac{V_2}{a^*_{5,4,u}} - {b'}^*_{5,4} \frac{a_{5,4,d}}{a^*_{5,4,u}} =
\frac{\lambda_c}{a^*_{5,7,u}} - {b'}^*_{5,7} \frac{a_{5,4,d}}{a^*_{5,7,u}},
\nonumber\\
&& \frac{V_1}{a^*_{5,7,u}} - {b'}^*_{5,7} \frac{a_{5,7,d}}{a^*_{5,7,u}} =
-\frac{\lambda_c}{a^*_{5,4,u}} - {b'}^*_{5,4} \frac{a_{5,7,d}}{a^*_{5,4,u}},
\label{C10}
\end{eqnarray}
which gives
\begin{eqnarray} 
&&{b'}^*_{5,4} \frac{1}{a^*_{5,4,u}} - {b'}^*_{5,7} \frac{1}{a^*_{5,7,u}} =
\frac{V_2}{a^*_{5,4,u} a_{5,4,d}} - \frac{\lambda_c}{a^*_{5,7,u} a_{5,4,d}}, 
\nonumber\\
&&{b'}^*_{5,4} \frac{1}{a^*_{5,4,u}} - {b'}^*_{5,7} \frac{1}{a^*_{5,7,u}} =
-\frac{V_1}{a^*_{5,7,u} a_{5,7,d}} - \frac{\lambda_c}{a^*_{5,4,u} a_{5,7,d}}.   
\label{C11}
\end{eqnarray}
Since both left sides are the same, the two right sides must be equal,
so one has
\begin{eqnarray}
V_2 + V_1 yz =\lambda_c(z-y), \quad z= \frac{a^*_{5,4,u}}{a^*_{5,7,u}}, \:
y= \frac{a_{5,4,d}}{a_{5,7,d}}.  
\label{C12}  
\end{eqnarray}
Then one uses further only the first equality of (\ref{C11}) which provides
the first row below
\begin{eqnarray}
&&{b'}^*_{5,4}={b'}^*_{5,7} \frac{a^*_{5,4,u}}{a^*_{5,7,u}} +
\frac{V_2}{a_{5,4,d}} - \frac{\lambda_c}{a_{5,4,d}} \frac{a^*_{5,4,u}}{
  a^*_{5,7,u}}
\nonumber\\
&&{b'}^*_{5,4} = \frac{t_c - a_{5,7,u}a^*_{5,4,u}}{b'_{5,7}} =
\frac{t_c - z |a_{5,7,u}|^2}{b'_{5,7}}
\label{C13}
\end{eqnarray}
where the second line is taken from the first $t_c$ relation of the last
row of Eqs.(\ref{C8}). From the equality of the two terms from
(\ref{C13}), one finds
\begin{eqnarray}
t_c - z  |a_{5,7,u}|^2 = |b'_{5,7}|^2 z + \frac{b'_{5,7}}{a_{5,4,d}}
(V_2 - z \lambda_c)
\label{C14}  
\end{eqnarray}
At first view it seems that (\ref{C14}) provides a quadratic equation for
$b'_{5,7}$, but is not so, since from the first line of (\ref{C8}) one has
$|a_{5,7,u}|^2=I_{1,u}-|b'_{5,7}|^2$, hence $|b'_{5,7}|^2$ cancels out from
(\ref{C14}), and we find
\begin{eqnarray}
b'_{5,7} = W_1 a_{5,4,d}, \quad W_1 = \frac{t_c-zI_{1,u}}{V_2-z\lambda_c}.
\label{C15}  
\end{eqnarray}
The other relations from (\ref{E22},\ref{E23})can be deduced in similar manner.
Indeed, expressing not $b_{5,4},b_{5,7}$ but $b'_{5,4}, b'_{5,7}$ from (\ref{C9}),
and using again the first $t_c$ expression from the last row of (\ref{C8}), we
find $b'_{5,4}=W_2 a_{5,4,d}$, and similarly, using the second $t_c$ expression
from the last row of (\ref{C8}) we obtain $b^*_{5,7}=W_3 a^*_{5,4,u}$ and
$b^*_{5,4}=W_4 a^*_{5,4,u}$.

Now I explain how (\ref{E24}) is obtained, by exemplifying the deduction
procedure of the first equality of (\ref{E24}): Deducing
$b'_{5,7} = W_1 a_{5,4,d}$ and $b'_{5,4}=W_2 a_{5,4,d}$, one expresses
$b'_{5,4}/b'_{5,7} = W_2/W_1$. Now it must be taken into account that from the
third line of (\ref{C8}) we also have a such ratio, namely
${b'}^*_{5,4}/{b'}^*_{5,7}= (a_{5,7,d}/a_{5,4,d})(V_2-a^*_{5,4,u}b_{5,4})/(V_1-a^*_{5,7,u}
b_{5,7})$. Equating these two ratios one obtain
\begin{eqnarray}
&&\big(\frac{W_2}{W_1}\big)^* = \frac{V_2 - a^*_{5,4,u}b_{5,4}}{V_1 - a^*_{5,7,u}
b_{5,7}}  \big( \frac{a_{5,7,d}}{a_{5,4,d}} \big) = \frac{1}{y}
\frac{V_2 -|a_{5,4,u}|^2 (b_{5,4}/a_{5,4,u})}{V_1 -a^*_{5,7,u}a_{5,4,u}
  (b_{5,7}/a_{5,4,u})}
\nonumber\\
&&=\frac{1}{y}\frac{V_2 -|a_{5,4,u}|^2 W_4^*}{V_1 -(1/z)|a_{5,4,u}|^2W_3^*}, 
\label{C16}
\end{eqnarray}  
from where, taking into account that $V_1,V_2$ are real, one finds
\begin{eqnarray}
\frac{W_2}{W_1} = \frac{1}{y^*} \frac{V_2 -|a_{5,4,u}|^2 W_4}{[V_1 -
(1/z^*)|a_{5,4,u}|^2W_3]},
\label{C17}
\end{eqnarray}
Taking into account that $y,z$ have the same phase factor, one obtains from
(\ref{C17}) the first equality of (\ref{E24}). By starting from the
$b_{5,4}/b_{5,7}$ ratio, the second equality of (\ref{E24}) can be similarly
deduced.

\section{Deduction of $\hat X^{\dagger}_i$ at ${\bf B}=0$}

Here  we solve the system of equations Eq.(\ref{E28}). The last four equations,
as explained, provide $x^*_{1,\uparrow} = x^*_{7',\uparrow} = y^*_{1,\downarrow} =
y^*_{7',\downarrow} = 0$, hence one remains with 20 unknown parameters and 18
equations.

One starts with the equations containing two terms, namely, the 7th,8th
equalities and 17th,18th equalities which provide
\begin{eqnarray}
x^*_{6,\uparrow}= - \frac{t_f}{\epsilon_4} x^*_{5,\uparrow}, \:
x^*_{6',\uparrow}= - \frac{t_f}{\epsilon_4} x^*_{5',\uparrow}, \: 
y^*_{6,\downarrow}= - \frac{t_f}{\epsilon_4} y^*_{5,\downarrow}, \:
y^*_{6',\uparrow}= - \frac{t_f}{\epsilon_4} y^*_{5',\uparrow}.
\label{D1}
\end{eqnarray}
Now from the first, second and 15th,16th equalities one finds
\begin{eqnarray}
x^*_{5,\uparrow} = - x^*_{2,\uparrow}, \: x^*_{5',\uparrow} = - x^*_{3',\uparrow}, \: 
y^*_{5,\downarrow} = - y^*_{2,\downarrow}, \: y^*_{5',\uparrow} = - y^*_{3',\downarrow}.
\label{D2}
\end{eqnarray}
After this step, introducing the notation $A=(\epsilon^2_2-t^2_h)/[t_h(t^2 +
\lambda^2)]$, from the fifth and sixth equations we obtain  
\begin{eqnarray}
x^*_{4,\uparrow} = A (t x^*_{2,\uparrow} - \lambda  y^*_{2,\downarrow}), \:
y^*_{4,\downarrow} = A (\lambda x^*_{2,\uparrow} + t y^*_{2,\downarrow}),  
\label{D3}  
\end{eqnarray}
and similarly, from the 11th and 12th equations one finds
\begin{eqnarray}
x^*_{4',\uparrow} = A (t x^*_{3',\uparrow} - \lambda  y^*_{3',\downarrow}), \:
y^*_{4',\downarrow} = A (\lambda x^*_{3',\uparrow} + t y^*_{3',\downarrow}).  
\label{D4}  
\end{eqnarray}
At this moment the relations presented in (\ref{E29}) have been obtained.
The remaining equations are the 9th and 10th equalities, which become of the
form
\begin{eqnarray}
&&x^*_{4,\uparrow} a_{5,4} + x^*_{4',\uparrow} a_{5,7} + y^*_{4,\downarrow} b_{5,4} +  
y^*_{4',\downarrow} b_{5,7} = 0, 
\nonumber\\
&&-x^*_{4,\uparrow} b_{5,4} - x^*_{4',\uparrow} b_{5,7} + y^*_{4,\downarrow} a_{5,4} +  
y^*_{4',\downarrow} a_{5,7} = 0,
\label{D5}
\end{eqnarray}
where introducing (\ref{D3},\ref{D4}), one obtains
\begin{eqnarray}
&&\alpha x^*_{2,\uparrow} + \gamma y^*_{2,\downarrow} = X = -\beta x^*_{3',\uparrow}
-\delta y^*_{3',\downarrow},  
\nonumber\\
&&-\gamma x^*_{2,\uparrow} + \alpha y^*_{2,\downarrow} = Y = \delta x^*_{3',\uparrow}
-\beta y^*_{3',\downarrow},  
\label{D6}
\end{eqnarray}
where $\alpha,\beta,\gamma,\delta$ are defined  below (\ref{E30}). From
(\ref{D6}) one finds
\begin{eqnarray}
x^*_{2,\uparrow} = \frac{\alpha X - \gamma Y}{\alpha^2 + \gamma^2}, \quad
y^*_{2,\downarrow} = \frac{\gamma X + \alpha Y}{\alpha^2 + \gamma^2}.  
\label{D7}
\end{eqnarray}
Eq.(\ref{D7}) is presented in (\ref{E30}).

\section{Deduction of $\hat X^{\dagger}_i$ at ${\bf B} \ne 0$}

We solve now the system of equations Eq.(\ref{E28}) at ${\bf B} \ne 0$ in
conditions in which  Eq.(\ref{E31}) is satisfied.

From the 7th,8th equalities and 17th,18th equalities of (\ref{E28})
on finds
\begin{eqnarray}
x^*_{6,\uparrow}=- \frac{a_{4,5,u}}{a_{4,6,u}} x^*_{5,\uparrow}, \:
y^*_{6,\downarrow}=- \frac{a_{4,5,d}}{a_{4,6,d}} y^*_{5,\downarrow}, \:  
x^*_{6',\uparrow}=- \frac{a_{4,5,u}}{a_{4,6,u}} x^*_{5',\uparrow}, \:
y^*_{6',\downarrow}=- \frac{a_{4,5,d}}{a_{4,6,d}} y^*_{5',\downarrow},  
\label{EE1}
\end{eqnarray}
while from the first, second and 15th,16th equalities one obtains
\begin{eqnarray}
x^*_{5,\uparrow}=- \frac{a_{1,2,u}}{a_{1,5,u}} x^*_{2,\uparrow}, \:
y^*_{5,\downarrow}=- \frac{a_{1,2,d}}{a_{1,5,d}} y^*_{2,\downarrow}, \:  
x^*_{5',\uparrow}=- \frac{a_{3,3,u}}{a_{3,5,u}} x^*_{3',\uparrow}, \:
y^*_{5',\downarrow}=- \frac{a_{3,3,d}}{a_{3,5,d}} y^*_{5',\downarrow}.  
\label{EE2}
\end{eqnarray}
Using these results in the third and fourth equations, and in the 13th and 14th
equations of (\ref{E28}) one finds
\begin{eqnarray}
&&x^*_{3,\uparrow}= A_{3,\uparrow} x^*_{2,\uparrow}, \:
y^*_{3,\downarrow}= B_{3,\downarrow} y^*_{2,\downarrow}, \:
x^*_{2',\uparrow}= A_{2',\uparrow} x^*_{3',\uparrow}, \:
y^*_{2',\downarrow}= B_{2',\downarrow} y^*_{3',\downarrow},
\nonumber\\
&&A_{3,\uparrow}=\frac{1}{a_{2,3,u}}(\frac{a_{1,2,u}a_{2,5,u}}{a_{1,5,u}}-a_{2,2,u}) =-
e^{i\phi_1}\frac{\epsilon_2 -h}{t_h},
\label{EE3}\\
&&B_{3,\downarrow}=\frac{1}{a_{2,3,d}}(\frac{a_{1,2,d}a_{2,5,d}}{a_{1,5,d}}-a_{2,2,d}) =-
e^{i\phi_1}\frac{\epsilon_2 +h}{t_h},
\nonumber\\
&&A_{2',\uparrow}=\frac{1}{a_{2,2,u}}(\frac{a_{3,3,u}a_{2,5,u}}{a_{3,5,u}}-a_{2,3,u}) =-
e^{i(2\phi_2 +2\phi_3)}\frac{(\epsilon_2 - h)(\epsilon_3-h) -t^2_h +T_h(T_h +
\epsilon_3-h)}{T_h (\epsilon_3-h) + t^2_h},
\nonumber\\
&&B_{2',\downarrow}=\frac{1}{a_{2,2,d}}(\frac{a_{3,3,d}a_{2,5,d}}{a_{3,5,d}}-a_{2,3,d}) =-
e^{i(2\phi_2 +2\phi_3)}\frac{(\epsilon_2 + h)(\epsilon_3+h) -t^2_h +T_h(T_h +
\epsilon_3+h)}{T_h (\epsilon_3+h) + t^2_h}.
\nonumber
\end{eqnarray}
Using the obtained results in the fifth and sixth equations, and in the
11th and 12th equations of (\ref{E28}) one obtains
\begin{eqnarray}
&&x^*_{4,\uparrow}= \frac{{\bar Y}b_{3,4} y^*_{2,\downarrow} - {\bar X}a_{3,4,d}x^*_{2,
\uparrow}}{b_{3,4}b'_{3,4}-a_{3,4,d}a_{3,4,u}}, \:
y^*_{4,\downarrow}= \frac{{\bar X}b'_{3,4} x^*_{2,\uparrow} - {\bar Y}a_{3,4,u}y^*_{2,
\downarrow}}{b_{3,4}b'_{3,4}-a_{3,4,d}a_{3,4,u}},
\nonumber\\
&&x^*_{4',\uparrow}= \frac{{\bar V}b_{1,1} y^*_{3',\downarrow} - {\bar Z}a_{1,1,d}x^*_{
3',\uparrow}}{b_{1,1}b'_{1,1}-a_{1,1,d}a_{1,1,u}}, \:
y^*_{4',\downarrow}= \frac{{\bar Z}b'_{1,1} x^*_{3',\uparrow} - {\bar V}a_{1,1,u}y^*_{
3',\downarrow}}{b_{1,1}b'_{1,1}-a_{1,1,d}a_{1,1,u}},  
\label{EE4}
\end{eqnarray}
where
\begin{eqnarray}
&&{\bar X}= \frac{a_{1,2,u}a_{3,5,u}}{a_{1,5,u}}-a_{3,3,u}A_{3,\uparrow} = a_{3,3,u}
e^{-i(2\phi_2 +2\phi_3)} [1 + \frac{\epsilon_2 -h}{T_h}],
\nonumber\\  
&&{\bar Y}= \frac{a_{1,2,d}a_{3,5,d}}{a_{1,5,d}}-a_{3,3,d}B_{3,\downarrow} = a_{3,3,d}
e^{-i(2\phi_2 +2\phi_3)} [1 + \frac{\epsilon_2 +h}{T_h}], \:
\nonumber\\
&&{\bar Z}= \frac{a_{1,5,u}a_{3,3,u}}{a_{3,5,u}}-a_{1,2,u}A_{2',\uparrow} = a_{1,2,u}
e^{i(2\phi_2 +2\phi_3)} \frac{T^2_h+(\epsilon_2-h)(\epsilon_3-h) +2T_h(\epsilon_3-h)}{
T_h(\epsilon_3-h) +t^2_h}, 
\nonumber\\
&&{\bar V}= \frac{a_{1,5,d}a_{3,3,d}}{a_{3,5,d}}-a_{1,2,d}B_{2',\downarrow} = a_{1,2,d}
e^{i(2\phi_2 +2\phi_3)}\frac{T^2_h+(\epsilon_2+h)(\epsilon_3+h) +2T_h(\epsilon_3+h)}{
T_h(\epsilon_3+h) +t^2_h}.
\label{EE5}
\end{eqnarray}
From symmetry considerations taking $\epsilon_2=\epsilon_3$, from the deduced
result, excepting the last row, we obtain the results presented in (\ref{E32}).
In order to obtain the last row of (\ref{E32}), one uses the 9th and 10th
equalities of (\ref{E28}) obtaining
\begin{eqnarray}
&&\alpha' x^*_{2,\uparrow} + \beta' y^*_{2,\downarrow} = X',
\nonumber\\
&&\gamma' x^*_{2,\uparrow} + \delta' y^*_{2,\downarrow} = Y',
\label{EE6}
\end{eqnarray}
where the prefactors, and the right side terms are presented in (\ref{E34}).
From (\ref{EE6}) one obtains the first line of (\ref{E34}), hence all
coefficients of $\hat X^{\dagger}_i$ have been deduced.

\section{Other specific solutions for $\hat X^{\dagger}_i$.}

\subsection{Solution  when (\ref{E31}) is not satisfied.}

The present case is deduced at ${\bf B} \ne 0$, considering
\begin{eqnarray}
a_{5,7,d} a_{5,7,u} = b_{5,7} b'_{5,7}, \quad a_{5,4,d} a_{5,4,u} = b_{5,4} b'_{5,4}.
\label{F1}  
\end{eqnarray}
Following the notations from (\ref{E22}-\ref{E24}), the equalities (\ref{F1})
are satisfied at
\begin{eqnarray}
W_2 W_4 =1, \quad W_1W_3 = \frac{1}{yz}.
\label{F2}
\end{eqnarray}
The deduction technique at start follows the steps presented in Appendix E
up to Eq.(\ref{EE5}), and provides
\begin{eqnarray}
&&x^*_{1,\uparrow}= -\frac{b_{5,7}}{a_{5,7,u}} y^*_{1,\downarrow}, \:
x^*_{7',\uparrow}= -\frac{b_{5,4}}{a_{5,4,u}} y^*_{7',\downarrow}, \:
x^*_{6,\uparrow}= -\frac{a_{4,5,u}}{a_{5,6,u}} \frac{(M_{1,\uparrow}- a_{1,2,u}
x^*_{2,\uparrow})}{a_{1,5,u}}, \:
y^*_{6,\downarrow}= -\frac{a_{4,5,d}}{a_{5,6,d}} \frac{(M_{1,\downarrow}- a_{1,2,d}
y^*_{2,\downarrow})}{a_{1,5,d}}, 
\nonumber\\
&&x^*_{6',\uparrow}= -\frac{a_{4,5,u}}{a_{4,6,u}} \frac{(M_{2,\uparrow}- a_{3,3,u}
x^*_{3',\uparrow})}{a_{3,5,u}}, \:
y^*_{6',\downarrow}= -\frac{a_{4,5,d}}{a_{5,6,d}} \frac{(M_{2,\downarrow}- a_{3,3,d}
y^*_{3',\downarrow})}{a_{3,5,d}}, \: x^*_{5,\uparrow}=\frac{(M_{1,\uparrow}- a_{1,2,u}
x^*_{2,\uparrow})}{a_{1,5,u}},
\nonumber\\
&&y^*_{5,\downarrow}=\frac{(M_{1,\downarrow}- a_{1,2,d} y^*_{2,\downarrow})}{a_{1,5,d}}, \:
x^*_{5',\uparrow}=\frac{(M_{2,\uparrow}- a_{3,3,u} x^*_{3',\uparrow})}{a_{3,5,u}}, \:
y^*_{5',\downarrow}=\frac{(M_{2,\downarrow}- a_{3,3,d} y^*_{3',\downarrow})}{a_{3,5,d}},
\nonumber\\
&&x^*_{3,\uparrow}= A^0_{3,\uparrow} + A_{3,\uparrow} x^*_{2,\uparrow}, \:
y^*_{3,\downarrow}= B^0_{3,\downarrow} + B_{3,\downarrow} y^*_{2,\downarrow}, \:
x^*_{2',\uparrow}= A^0_{2',\uparrow} + A_{2',\uparrow} x^*_{3',\uparrow}, \:
y^*_{2',\downarrow}= B^0_{2',\downarrow} + B_{2',\uparrow} y^*_{3',\downarrow}, 
\label{F3}
\end{eqnarray}    
where $A_{3,\uparrow},B_{3,\downarrow},A_{2',\uparrow},B_{2',\downarrow}$ are given in 
(\ref{EE3}), $y^*_{1,\downarrow}, y^*_{7',\downarrow}$ are arbitrary, and
\begin{eqnarray}
&&M_{1,\uparrow}=-a_{1,1,u}x^*_{1,\uparrow} -b_{1,1} y^*_{1,\downarrow}, \:
M_{1,\downarrow}=-a_{1,1,d}y^*_{1,\downarrow} -b'_{1,1} x^*_{1,\uparrow}, \:
M_{2,\uparrow}=-a_{3,4,u}x^*_{7',\uparrow} -b_{3,4} y^*_{7',\downarrow},
\nonumber\\
&&M_{2,\downarrow}=-b'_{3,4}x^*_{7',\uparrow} -a_{3,4,d} y^*_{7',\downarrow}, \:
A^0_{3,\uparrow}=-\frac{a_{2,5,u}M_{1,\uparrow}}{a_{2,3,u}a_{1,5,u}}, \:
B^0_{3,\downarrow}=-\frac{a_{2,5,d}M_{1,\downarrow}}{a_{2,3,d}a_{1,5,d}},
\nonumber\\
&&A^0_{2',\uparrow}=-\frac{a_{2,5,u}M_{2,\uparrow}}{a_{2,2,u}a_{3,5,u}}, \:
B^0_{2',\downarrow}=-\frac{a_{2,5,d}M_{2,\downarrow}}{a_{2,2,d}a_{3,5,d}}. 
\label{F4}
\end{eqnarray}
Furthermore, one obtains
\begin{eqnarray}
&&x^*_{4,\uparrow} =\frac{({\bar Y}_0 b_{3,4}-{\bar X}_0 a_{3,4,d}) + ({\bar Y}
b_{3,4} y^*_{2,\downarrow} - {\bar X} a_{3,4,d} x^*_{2,\uparrow})}{b'_{3,4}b_{3,4} - 
a_{3,4,d}a_{3,4,u}},
\nonumber\\
&&y^*_{4,\downarrow} =\frac{({\bar X}_0 b'_{3,4}-{\bar Y}_0 a_{3,4,u}) + ({\bar X}
b'_{3,4} x^*_{2,\uparrow} - {\bar Y} a_{3,4,u} y^*_{2,\downarrow})}{b'_{3,4}b_{3,4} - 
a_{3,4,d}a_{3,4,u}},
\nonumber\\
&&x^*_{4',\uparrow} =\frac{({\bar V}_0 b_{1,1}-{\bar Z}_0 a_{1,1,d}) + ({\bar V}
b_{1,1} y^*_{3',\downarrow} - {\bar Z} a_{1,1,d} x^*_{3',\uparrow})}{b'_{1,1}b_{1,1} - 
a_{1,1,d}a_{1,1,u}},
\nonumber\\
&&y^*_{4',\downarrow} =\frac{({\bar Z}_0 b'_{1,1}-{\bar V}_0 a_{1,1,u}) + ({\bar Z}
b'_{1,1} x^*_{3',\uparrow} - {\bar V} a_{1,1,u} y^*_{3',\downarrow})}{b'_{1,1}b_{1,1} - 
a_{1,1,d}a_{1,1,u}},  
\label{F5}
\end{eqnarray}
where ${\bar X},{\bar Y},{\bar Z},{\bar V}$ are given in (\ref{EE5}), while the
remaining parameters are defined as
\begin{eqnarray}
&&{\bar X}_0= -M_{1,\uparrow} (\frac{a_{3,5,u}}{a_{1,5,u}} - \frac{a_{3,3,u} a_{2,5,u}}{
a_{2,3,u} a_{1,5,u}}, \:
{\bar Y}_0= -M_{1,\downarrow} (\frac{a_{3,5,d}}{a_{1,5,d}} - \frac{a_{3,3,d} a_{2,5,d}}{
a_{2,3,d} a_{1,5,d}},
\nonumber\\
&&{\bar Z}_0= -M_{2,\uparrow} (\frac{a_{1,5,u}}{a_{3,5,u}} - \frac{a_{1,2,u} a_{2,5,u}}{
a_{2,2,u} a_{3,5,u}}, \:
{\bar V}_0= -M_{2,\downarrow} (\frac{a_{1,5,d}}{a_{3,5,d}} - \frac{a_{1,2,d} a_{2,5,d}}{
a_{2,2,d} a_{3,5,d}}.  
\label{F6}
\end{eqnarray}
Starting from this point, because of (\ref{F1}), the deduction procedure
strongly differs from that applied in Appendix E. The remaining equations from 
(\ref{E28}) are
\begin{eqnarray}
&&x^*_{4,\uparrow} a_{5,4,u} + y^*_{4,\downarrow} b_{5,4} + x^*_{4',\uparrow} a_{5,7,u} +
y^*_{4',\downarrow} b_{5,7} = 0,
\nonumber\\  
&&x^*_{4,\uparrow} b'_{5,4} + y^*_{4,\downarrow} a_{5,4,d} + x^*_{4',\uparrow} b'_{5,7} +
y^*_{4',\downarrow} a_{5,7,d} = 0.
\label{F7}  
\end{eqnarray}
Using (\ref{F1}), the system of equations (\ref{F7}) becomes of the form
\begin{eqnarray}
&&a_{5,4,u}b'_{5,7} (x^*_{4,\uparrow}b'_{5,4} + y^*_{4,\downarrow}a_{5,4,d}) +
a_{5,7,u}b'_{5,4} (x^*_{4',\uparrow}b'_{5,7} + y^*_{4',\downarrow}a_{5,7,d}) = 0,
\nonumber\\
&&(x^*_{4,\uparrow}b'_{5,4} + y^*_{4,\downarrow}a_{5,4,d}) +
(x^*_{4',\uparrow}b'_{5,7} + y^*_{4',\downarrow}a_{5,7,d}) = 0,  
\label{F8}
\end{eqnarray}
from where it results
\begin{eqnarray}
(x^*_{4,\uparrow}b'_{5,4} + y^*_{4,\downarrow}a_{5,4,d}) = 0, \:
(x^*_{4',\uparrow}b'_{5,7} + y^*_{4',\downarrow}a_{5,7,d}) = 0.
\label{F9}
\end{eqnarray}
From here, using (\ref{F5}) one finds
\begin{eqnarray}
&&y^*_{2,\downarrow} = K_2 x^*_{2,\uparrow} + C_2, \:
y^*_{3',\downarrow} = K_3 x^*_{3',\uparrow} + C_3, \: x^*_{2,\uparrow} = p, \:
x^*_{3',\uparrow} = q,  
\nonumber\\
&&K_2 = \frac{{\bar X} (b'_{5,4}a_{3,4,d} - b'_{3,4} a_{5,4,d})}{{\bar Y}(b'_{5,4}
b_{3,4}-a_{3,4,u}a_{5,4,d})}, \:
C_2 = \frac{b'_{5,4}({\bar X}_0 a_{3,4,d}-{\bar Y}_0 b_{3,4})- a_{5,4,d}({\bar X}_0
b'_{3,4} - {\bar Y}_0a_{3,4,u})}{{\bar Y}(b'_{5,4}b_{3,4}-a_{3,4,u}a_{5,4,d})},  
\nonumber\\
&&K_3 = \frac{{\bar Z} (b'_{5,7}a_{1,1,d} - b'_{1,1} a_{5,7,d})}{{\bar V}(b'_{5,7}
b_{1,1}-a_{1,1,u}a_{5,7,d})}, \:
C_3 = \frac{b'_{5,7}({\bar Z}_0 a_{1,1,d}-{\bar V}_0 b_{1,1})- a_{5,7,d}({\bar Z}_0
b'_{1,1} - {\bar V}_0a_{1,1,u})}{{\bar V}(b'_{5,7}b_{1,1}-a_{1,1,u}a_{5,7,d})}, 
\label{F10}
\end{eqnarray}
where $p$, and $q$ are arbitrary.

\subsection{Solution on restricted number of sites.}

In this subsection one presents a specif ic solution which appears on the sites
2,3,5,6 inside a single cell (see Fig.2). The $\hat X^{\dagger}_i$ operator has in
the present case the expression
\begin{eqnarray}
\hat X^{\dagger}_{i,\sigma} = x^*_{2,\sigma} \hat c^{\dagger}_{i,2,\sigma} +
x^*_{3,\sigma} \hat c^{\dagger}_{i,3,\sigma} + x^*_{5,\sigma} \hat c^{\dagger}_{i,5,\sigma} +
x^*_{6,\sigma} \hat c^{\dagger}_{i,6,\sigma},
\label{F11}
\end{eqnarray}
where $\sigma$ is fixed. The equations (\ref{E25}) becomes in this case of the
form (for exemplification, one takes $\sigma=\uparrow$)
\begin{eqnarray}
&&x^*_{2,\uparrow} a_{1,2,u} + x^*_{5,\uparrow} a_{1,5,u} = 0,
\nonumber\\
&&x^*_{2,\uparrow} a_{2,2,u} + x^*_{3,\uparrow} a_{2,3,u} + x^*_{5,\uparrow} a_{2,5,u} = 0,
\nonumber\\
&&x^*_{3,\uparrow} a_{3,3,u} + x^*_{5,\uparrow} a_{3,5,u} = 0,
\nonumber\\
&&x^*_{5,\uparrow} a_{4,5,u} + x^*_{6,\uparrow} a_{4,6,u} = 0.
\label{F12}
\end{eqnarray}
The solution of the matching equation (\ref{F12}) becomes of the form
\begin{eqnarray}
x^*_{2,\uparrow} = - \frac{a_{1,5,u}}{a_{1,2,u}} x^*_{5,\uparrow}, \:
x^*_{3,\uparrow} = - \frac{a_{3,5,u}}{a_{3,3,u}} x^*_{5,\uparrow}, \:
x^*_{6,\uparrow} = - \frac{a_{4,5,u}}{a_{4,6,u}} x^*_{5,\uparrow}, 
\label{F13}  
\end{eqnarray}
where $x^*_{5,\uparrow}=p$ is an arbitrary parameter, but the condition
\begin{eqnarray}
a_{2,5,u} = a_{2,2,u} \frac{a_{1,5,u}}{a_{1,2,u}} +  a_{2,3,u} \frac{a_{3,5,u}}{a_{3,3,u}}
\label{F14}
\end{eqnarray}
must be satisfied. In terms of the Hamiltonian parameters this condition means
$t_h = -\epsilon_2$ at ${\bf B}=0$, and $T_h = -\epsilon_2 + h$ at ${\bf B} \ne
0$, ($\epsilon_3=\epsilon_2$ is considered from symmetry considerations).
Similar solution holds for $\sigma=\downarrow$ as well.


\end{document}